\newcommand {\mH}	{{\rm [m/H]}}	% Roman font metallicity [m/H]
\newcommand {\FeH} 	{{\rm [Fe/H]}}	% Roman font metallicity [Fe/H]
\newcommand {\OFe} 	{{\rm [O/Fe]}}	% Roman font metallicity [O/Fe]
\newcommand {\V}	{{\sl V\/}}	% V magnitude
\newcommand {\I}	{{\sl I\/}}	% I magnitude
\newcommand {\eg} 	{e.g.\ }	% e.g.
\newcommand {\etal}	{et al.\ }	% et al.
\newcommand {\fig}	{Fig.\ }	% Text for ``figure''
\newcommand {\figs}	{Figs.\ }	% Text for ``figures''
\newcommand {\sect}[1]	{Sec.\ #1}	% Text for ``section''

\documentstyle[11pt,aaspp4]{article}
\tighten
  
\received{}
\revised{}
\accepted{}
\journalid{}{}
\articleid{}{}
\paperid{}

\lefthead{Holland \etal}
\righthead{The Halo of M31}

\slugcomment{accepted for publication in the {\sl AJ}}

\begin{document}

\title{Deep {\sl HST\/} \V- and \I-Band Observations of the Halo of
M31: Evidence for Multiple Stellar Populations\footnote{Based in
observations with the NASA/ESA {\sl Hubble Space Telescope}, obtained
at the Space Telescope Science Institute, which is operated by the
Association of Universities for Research in Astronomy, Inc., under
NASA contract NAS5-26555.}}

\author{Stephen Holland, Gregory G. Fahlman, \& Harvey B. Richer}

\affil{Department of Physics \& Astronomy \\
\#129--2219 Main Mall \\
University of British Columbia \\
Vancouver, B.C., Canada \\
V6T 1Z4}

\begin{abstract}
	We present deep ($V \simeq 27$) \V- and \I-band stellar
photometry obtained using the {\sl Hubble Space Telescope\/}'s WFPC2
in two fields in the M31 halo.  These fields are located 32\arcmin~and
50\arcmin~from the center of M31 approximately along the SE minor axis
at the locations of the M31 globular star clusters G302 and G312
respectively.  The M31 halo luminosity functions are not consistent
with a single high-metallicity population but are consistent with a
mix of 50\% to 75\% metal-rich stars and 25\% to 50\% metal-poor
stars.  This is consistent with the observed red giant branch
morphology, the luminosity of the horizontal branch, and the presence
of RR Lyrae stars in the halo of M31.  The morphology of the red giant
branch indicates a spread in metallicity of $-2 \lesssim \mH \lesssim
-0.2$ with the majority of stars having $\mH \simeq -0.6$, making the
halo of M31 significantly more metal-rich than either the Galactic
halo or the M31 globular cluster system.  The horizontal branch is
dominated by a red clump similar to the horizontal branch of the
metal-rich Galactic globular cluster 47 Tuc but a small number of blue
horizontal branch stars are visible, supporting the conclusion that
there is a metal-poor component to the stellar population of the halo
of M31.  The number of blue horizontal stars is smaller than would be
expected from the observed metallicity distribution but it is not
clear if this is due to the photometric limits of our data or a second
parameter effect in the halo of M31.  We find a lower limit to the
helium abundance of $Y \simeq 0.20$ to $0.27$, comparable with the
Galactic value.  Luminosity functions show weak evidence that the
$R_{\rm M31} = 50$\arcmin~field contains a higher fraction of
metal-poor stars than the $R_{\rm M31} = 32$\arcmin~field but the
metallicity distributions of the RGB stars in each field strongly
suggest that both fields have the same mix of stellar populations.
\end{abstract}


%%%%%%%%%%%%%%%%%%%%%%%%%%%%%%%%%%%%%%%%%%%%%%%%%%%%%%%%%%%%%%%%%%%%%%%%

\section{Introduction\label{SECTION:intro}}

	The stellar populations in the halo of the Andromeda Galaxy
(M31 $=$ NGC 224) provide a direct tracer of the star formation
history, and the early evolution, of that galaxy.  M31 is
approximately twice as massive as, and slightly more metal-rich than,
the Milky Way.  It is our nearest large galactic neighbor and has a
similar Hubble type so M31 is generally regarded as being a twin of
our Galaxy.  The low inclination ($i = 12\fdg5$, Hodge
\markcite{H92}1992) makes it possible to study the stellar halo with
minimal contamination from disk stars.

	The earliest published color-magnitude diagrams (CMDs) of the
M31 halo were obtained by Crotts \markcite{C86}(1986) and Mould
\& Kristian \markcite{MK86}(1986).  Mould \& Kristian \markcite{MK86}(1986)
studied a field $\sim$7 kpc from the center of M31 along the SE minor
axis and found a mean metallicity of $\overline{\FeH} \sim -0.6$,
comparable to that of the metal-rich Galactic globular star cluster
(GC) 47 Tuc.  Subsequent ground-based studies by Pritchet \& van den
Bergh \markcite{PvdB88}(1988), Christian \& Heasley
\markcite{CH91}(1991), Davidge \markcite{Da93}(1993), Durrell
\etal \markcite{DH94}(1994), and Couture \etal \markcite{CR95}(1995)
produced CMDs reaching down to near the level of the horizontal branch
(HB) stars.  These studies confirmed that the M31 halo has a mean
metallicity similar to that of 47 Tuc, making the M31 halo
significantly more metal-rich than the Galactic halo.  Several of
these studies have shown that the observed spread in color across the
red giant branch (RGB) is larger than what would be expected from
photometric uncertainties and thus could be due to an intrinsic spread
in the metallicity of the halo of M31 ($0.3 \lesssim \sigma_\mH
\lesssim 0.5$).  A deep \V- and \I-band study of the halo of M31 in the
vicinity of five of M31's GCs by Couture \etal \markcite{CR95}(1995)
found that the halo was dominated by a stellar population with a mean
metallicity comparable to that of 47 Tuc but that a small component of
the RGB stars had $\FeH \simeq -1$ to $-1.5$.  This is consistent with
the presence of the RR Lyrae variables in the halo of M31 as observed
by Pritchet \& van den Bergh \markcite{PvdB87}(1987).  Recently Rich
\etal \markcite{RM96}(1996) used the {\sl Hubble Space Telescope\/}
({\sl HST\/}) to obtain CMDs and luminosity functions (LFs) of the
metal-rich M31 GC Mayall II ($=$ G1) and the field near it.  They
found that the halo LF was steeper than the G1 LF and had a less
pronounced HB.  They compared their M31 halo LF to LFs of several
Galactic GCs covering a range of metallicities and found that no
single-metallicity LF could reproduce the M31 halo LF.

	Deep CMD studies have the potential to provide direct
information on the chemical composition and age of the halo of M31.
This is important as there is some debate as to the ages of the GCs in
the M31 system.  There is considerable evidence that the halo of M31
is an old stellar system.  Firstly, integrated colors (Frogel \etal
\markcite{FP80}1980, Bohlin \etal \markcite{BD93}1993) and CMDs suggest
that the M31 GC system has an age comparable to that of the Galactic
GC system.  Secondly, no extended giant branch of AGB stars has been
detected in the halo or the bulge of M31 (Rich \& Mighell
\markcite{RM95}1995) as would be expected if an intermediate age
population was present.  Thirdly, the presence of RR Lyrae variables
implies that at least part of the stellar population is old.  On the
other hand, measurements of H$\beta$ enhancements (Burstein \etal
\markcite{BF84}1984, Tripicco \markcite{T89}1989) in spectra of M31
GCs relative to spectra of Galactic GCs have led to speculation that
the M31 GCs could contain a higher fraction of main-sequence stars
than the Galactic GCs do, implying that the GCs are younger than their
Galactic counterparts.  It has recently been suggested (\eg Faber
\markcite{F95}1995) that H$\beta$ enhancements in red elliptical
galaxies are due to the presence of a population of young
main-sequence turn-off stars.  It is, therefore, of great interest to
obtain CMD-based determinations of the mix of ages and metallicities
in the nearest population of stars resembling a ``red elliptical''
galaxy---the halo of M31.

	Ashman \& Bird \markcite{AB93}(1993) identified several
distinct groups of GCs in the M31 system which may be associated with
sub-structure in the stellar halo of M31.  This is similar to the
situation in the Galactic halo where several GCs are believed to have
originated in external galaxies that are orbiting, or have been
accreted by, the Galaxy (\eg Lin \& Richer \markcite{LR92}1992, Ibata
\etal \markcite{IG94}1994).  If accretion has played a significant
role in adding material to the halo of M31 then stars from different
accretion events could have different ages and metallicities resulting
in the M31 halo resembling a patchwork of different stellar
populations.  Any such clumps in the M31 halo will be seen in
projection against the rest of the halo so a CMD will show features
from both the underlying halo population and the population of the
clump.  If the halo of M31 is made up of accreted clumps then CMDs of
the halo will show evidence of multiple populations.  The exact mix of
stellar populations will vary from one accreted clump to the next.

	This paper presents deep ($V \simeq 27$) \V- and \I-band
photometry in two fields 32\arcmin~and 50\arcmin~from the center of
M31 approximately along the SE minor axis.  These fields are at
projected distances of 7.6 kpc and 10.8 kpc respectively from the
center of M31 assuming a distance modulus of $\mu = 24.3$ (van den
Bergh \markcite{vdB91}1991).  This location in the M31 halo is roughly
analogous to the distance from the center of the Milky Way to the Sun.

%%%%%%%%%%%%%%%%%%%%%%%%%%%%%%%%%%%%%%%%%%%%%%%%%%%%%%%%%%%%%%%%%%%%%%%%%

\section{Observations and Data Reductions\label{SECTION:obs+data}}

\subsection{Observations\label{SECTION:obs}}

	We obtained deep \V- and \I-band images of the halo of M31 in
the vicinity of two bright GCs, G302 ($\alpha_{2000.0} = 00^{\rm h}
45^{\rm m} 25\fs2,~ \delta_{2000.0} = +41\arcdeg 05\arcmin 30\arcsec$)
and G312 ($\alpha_{2000.0} = 00^{\rm h} 45^{\rm m} 58\fs8,~
\delta_{2000.0} = +40\arcdeg 42\arcmin 32\arcsec$).  This data was
obtained using the {\sl HST\/} as part of a project (cycle 5 program
\#5609) to study the internal structures and stellar populations of
GCs in M31 (Holland \etal \markcite{HF96}1996).  Table
\ref{TABLE:obs_log} lists the exposures obtained in each filter.
Exposure times were selected to avoid saturating the cores of the two
GCs being imaged.  The total exposure times were 4320 seconds in the
F555W (WFPC2 broadband \V) filter and 4160 seconds in the F814W (WFPC2
broadband \I) filter for each field.  The data was processed through
the standard STScI pipelines.  Known bad pixels were masked, and
geometric corrections were applied using standard techniques.

	The G302 field is located 32\arcmin~($R_{\rm M31} = 7.6$ kpc)
from the center of M31 approximately along the SE minor axis while the
G312 field is located 50\arcmin~($R_{\rm M31} = 10.8$ kpc) from the
center of M31 approximately along the SE minor axis.  Each GC was
centered on the WF3 CCD leaving the WF2 and WF4 CCDs to image the halo
of M31.  The tidal radii ($r_t$), determined by fitting Michie--King
models (Michie \markcite{M63}1963, King \markcite{K66}1966) to the
surface brightness profiles of G302 and G312, are $r_t
\simeq10$\arcsec~(Holland\markcite{H97} 1997).  Grillmair \etal
\markcite{GF95}(1995) found evidence for extra-tidal tails, extending
to $\gtrsim 2 r_t$ to $3 r_t$, in several Galactic GCs.  Holland
\markcite{H97}(1997) found weak evidence for stellar density
enhancements out to $\sim 2 r_t$ in some M31 GCs including G302 and
G312.  Since the GCs are located approximately 40\arcsec~($\sim 4
r_t$) from the edge of the WF3 CCD the WF2 and WF4 CCDs should contain
little, if any, contamination from an extended halo of GC stars.

\subsection{Data Reductions\label{SECTION:data}}

	Photometry was performed on stars imaged by the WFPC2's WF2
and WF4 CCDs using the {\sc DaoPhot II}/{\sc AllFrame} software
(Stetson \markcite{S87}1987, Stetson \markcite{S94}1994).  All
nineteen images for each field were re-registered to a common
coordinate system and median-combined.  This served to eliminate
cosmic ray hits and to increase the signal-to-noise ratio (S/N)
allowing fainter stars to be detected.  A square median-filter was run
over the combined image to determine large-scale gradients.  This
smoothed image was then subtracted from the combined image and a
constant sky added back to produce a median-subtracted image.  This
was done so that the star lists and photometry from the halo fields
would be obtained in exactly the same manner as the star lists and
photometry of the GC stars on the WF3 CCD (see Holland \etal
\markcite{HF96}1996).  The size of the square median-filter was
$\sim$5 times the stellar full-width at half-maximum (FWHM) to avoid
smoothing out structure in the stellar images.  The {\sc DaoPhot Find}
routine with a {\sc Find} threshold of $7.5$-$\sigma_{\rm sky}$ was
used to detect peaks on the median-subtracted image.  We experimented
with different {\sc Find} thresholds and found that $7.5\sigma_{\rm
sky}$ excluded most of the mis-identifications of noise spikes \&
cosmic ray events as stars at faint magnitudes while including almost
all of the stars down to near the photometric limit of the data.  The
resulting list of stellar candidates was used as input to {\sc
AllFrame}.  {\sc AllFrame} does simultaneous point-spread function
(PSF) fitting on all the {\sl original\/} images ({\sl not\/} the
combined median-subtracted image) to preserve the photometric
properties of the data.  The PSFs used (Stetson \markcite{S96}1996)
were Moffatians (Moffat \markcite{M69}1969) with $\beta = 1.5$ and a
look-up table of residuals.  The PSFs varied quadratically over each
WF CCD.  Only stars that appeared in at least 7 frames in each filter
were considered to be real stars.

\begin{figure}
\plotone{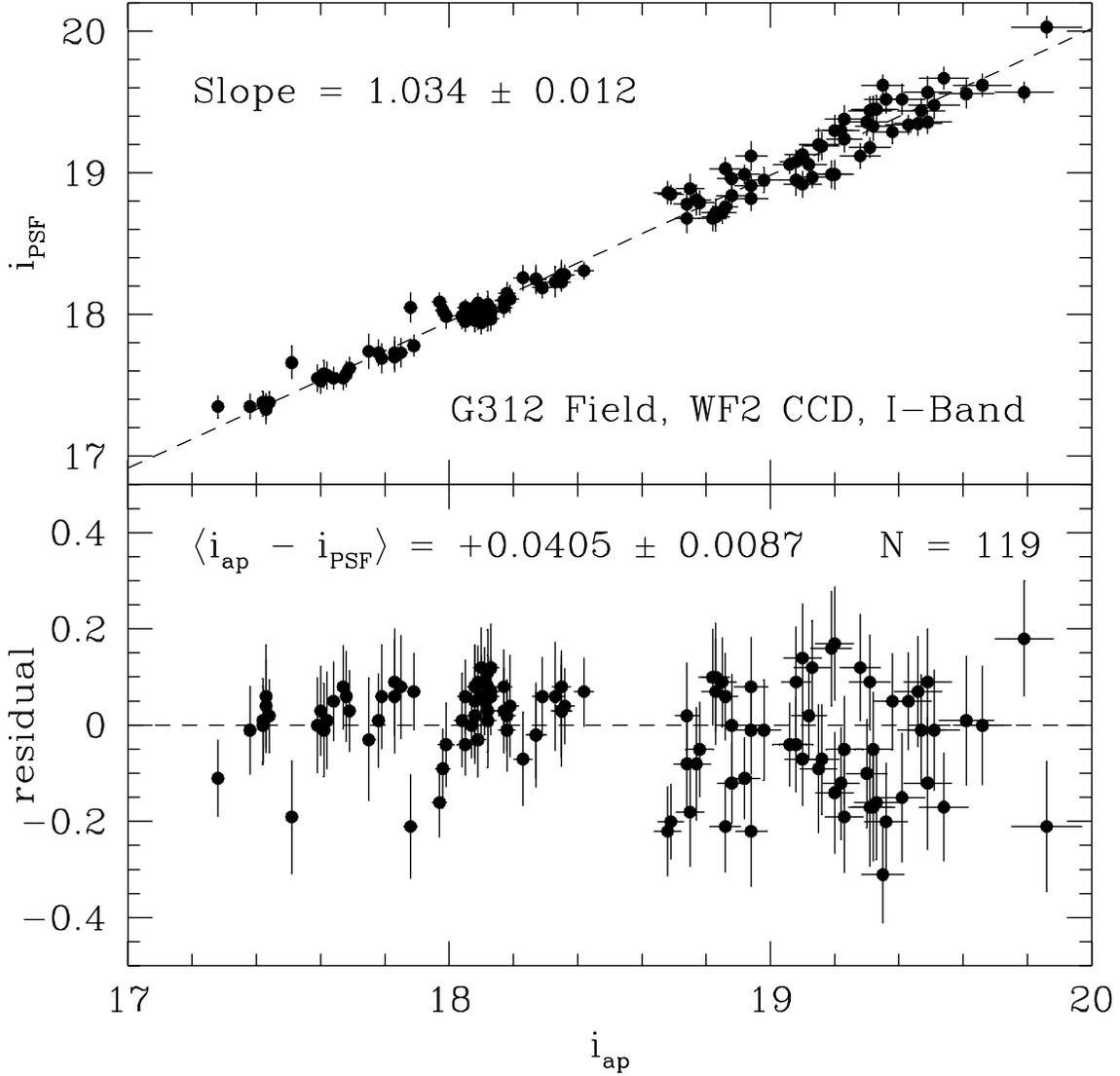}
\caption{The upper panel shows the aperture-magnitude vs PSF-magnitude
relation for the \I-band photometry from the WF2 CCD in the G302
field.  Stars were discarded until the slope of this relation
approached unity.  The lower panel show the residuals ($= (v_{\rm ap}
- v_{\rm PSF}) - \Delta v$, where $\Delta v = \langle v_{\rm ap} -
v_{\rm PSF} \rangle$).  Stars that failed the culling process
described in \sect{\ref{SECTION:data}} have not been
plotted. \label{FIGURE:apcor}}
\end{figure}

	Aperture corrections were obtained for each CCD in the
following manner.  First a set of $\sim$50 bright isolated stars was
selected on each of the WF2 and WF4 CCDs and all remaining stars were
subtracted from these images.  Next, the total magnitude in an
aperture with a 0\farcs5 radius (Holtzman \etal \markcite{HB95}1995b)
was measured for each star.  Aperture corrections were defined in the
sense $\Delta v = v_{\rm ap} - v_{\rm PSF}$ for each of the isolated
stars.  We discarded any star with a combined photometric uncertainty
of $\sqrt{\sigma^2_{\rm ap} + \sigma^2_{\rm PSF}} > 0.14$, which
corresponds to ${\rm S/N} \simeq 10$ for each of the aperture and PSF
magnitudes.  The size of the aperture correction should be independent
of magnitude so a plot of $v_{\rm ap}$ vs $v_{\rm PSF}$ (see \fig
\ref{FIGURE:apcor}) should yield a slope of unity with a zero-point
offset corresponding to the value of the aperture correction.
However, the WF CCDs suffer from considerable cosmic ray contamination
that may bias the observed aperture magnitudes towards brighter
magnitudes.  In addition the presence of background galaxies and
undetected faint stars can bias the $v_{\rm ap}$ -- $v_{\rm PSF}$
relation away from a slope of unity.  To correct for this we plotted
the PSF-magnitudes as a function of the aperture-magnitudes for each
WF CCD and iteratively discarded stars that biased the best-fitting
straight line away from a slope of unity until the slope approached
unity and discarding additional stars failed to drive the slope closer
to unity.  The aperture correction was then computed from the sample
of stars that survived this culling process.  Table \ref{TABLE:apcor}
lists the adopted aperture corrections for the WF2 and WF4 CCDs.  \fig
\ref{FIGURE:apcor} shows the residuals for the \I-band aperture
corrections to the WF2 CCD in the G312 field.

	The photometry was calibrated to standard Johnson--Cousins
\V-and \I-band magnitudes using the transformations of Holtzman
\etal \markcite{HB95}(1995b).  A charge transfer efficiency 
correction of 2\% (Holtzman \etal \markcite{HH95}1995a) was adopted
and a reddening of $E_{B-V} = 0.08 \pm 0.02$ (Burstein \& Heiles
\markcite{BH82}1982) was used.  We converted $E_{B-V}$ to $E_{V-I}$
using the relationship of Bessell \& Brett \markcite{BB88}(1988):
$E_{V-I} = 1.25 E_{B-V} = 0.10 \pm 0.03$, and adopted a distance
modulus of $\mu = (m-M)_0 = 24.3 \pm 0.1$ (van den Bergh
\markcite{vdb91}1991).  This value agrees well with other M31 distance
moduli obtained using Population II distance indicators.  Pritchet \&
van den Bergh \markcite{PvdB87}(1987) found $\mu = 24.34 \pm 0.15$
from RR Lyrae variables in the halo of M31 while Christian \& Heasley
\markcite{CH91}(1991) found $\mu \sim 24.3$ from the brightest red
giants in M31 GCs.  Studies of Population I distance indicators such
as Cepheids (\eg Freedman \& Madore \markcite{FM90}1990) and carbon
stars (\eg Brewer \etal \markcite{BR95}1995) find $\mu \simeq 24.4$,
slightly larger than the Population II value.  We have adopted the
lower distance modulus since the halo of M31 appears to consist of
Population II stars.

	Any stars for which {\sc AllFrame} returned a $\chi_{\rm DAO}
> 2$ were discarded.  The $\chi_{\rm DAO}$ value represents the ratio
of the observed pixel-to-pixel scatter in the fitting residuals to the
expected scatter.  Values that are significantly different from unity
indicate that an image was not well fit by the stellar PSF.  We used
an interactive approach to cull the star lists based on the
uncertainty in the calibrated magnitudes.  A plot of uncertainty vs
magnitude was made and a locus defined that corresponded to the
expected increase in photometric uncertainty as the stars became
fainter.  Any stars that were judged to have a significantly greater
uncertainty than the typical value for that magnitude was removed.
Visual examinations of the CMDs and the WF images suggested that we
did not remove legitimate stars from our sample.  We wish to emphasize
that this culling was done {\sl solely\/} on the basis of the observed
photometric uncertainty of each star, not on the star's location on
the CMD or on the CCD.  The photometric uncertainties for our culled
data are listed in Table
\ref{TABLE:errs}.

%%%%%%%%%%%%%%%%%%%%%%%%%%%%%%%%%%%%%%%%%%%%%%%%%%%%%%%%%%%%%%%%%%%%%%%

\section{The Color--Magnitude Diagrams\label{SECTION:cmds}}

\subsection{Contamination in the Field\label{SECTION:contam}}

\begin{figure}
\plotone{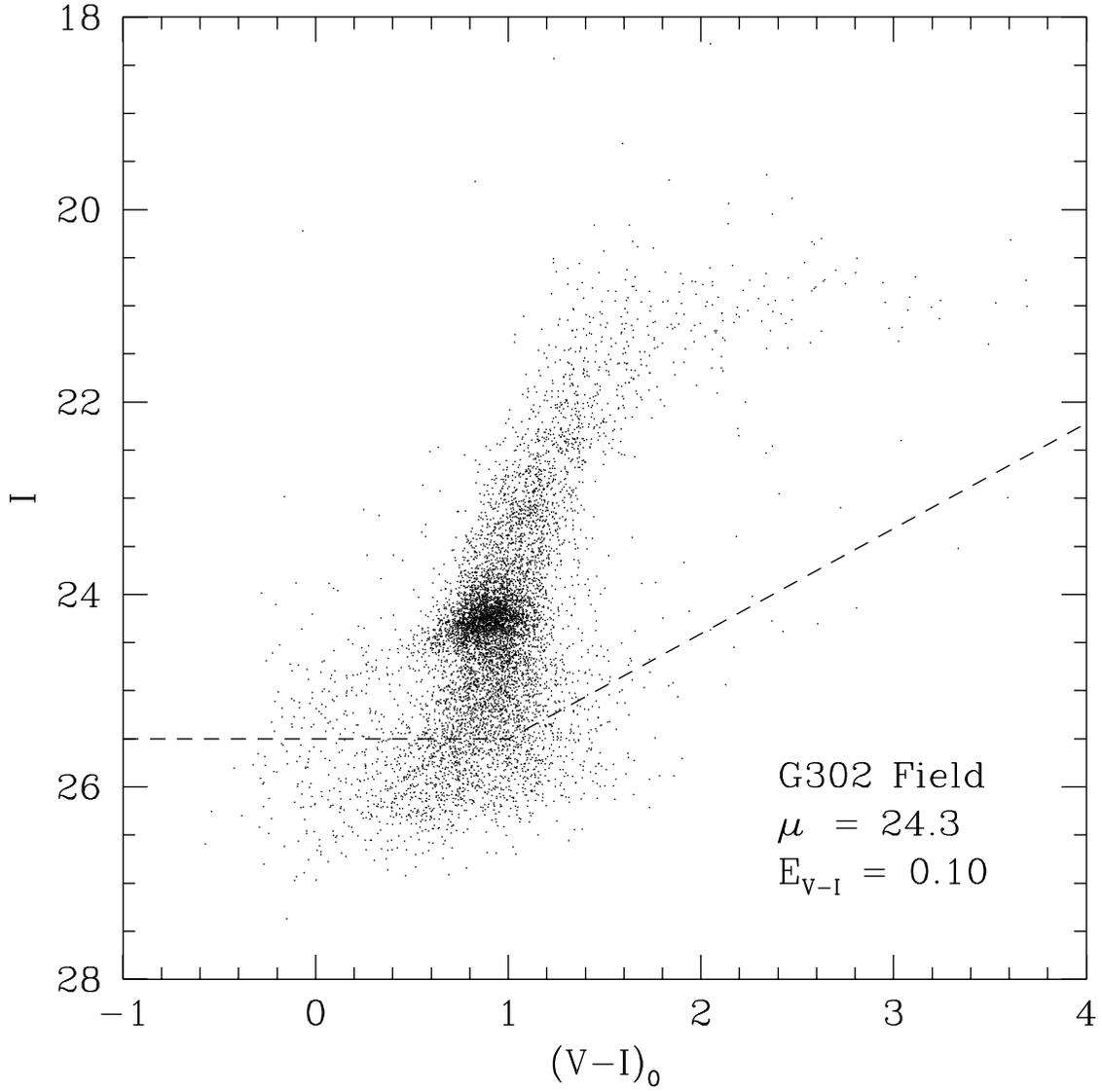}
\caption{This \fig shows the CMD for the M31 halo field around the GC
G302.  This diagram shows all star-like sources in our images.  No
attempt has been made to remove foreground or background
contamination.  The dashed line shows the location where the
photometric uncertainties are $\sigma_{(V-I)} \simeq 0.2$.  We believe
that this also corresponds to the approximate location where our data
becomes incomplete.  The distance modulus and extinction quoted are
the values that were used to calibrate our photometry (see Holtzman et
al. \protect\markcite{HB95}1995b). \label{FIGURE:G302_cmd}}
\end{figure}

\begin{figure}
\plotone{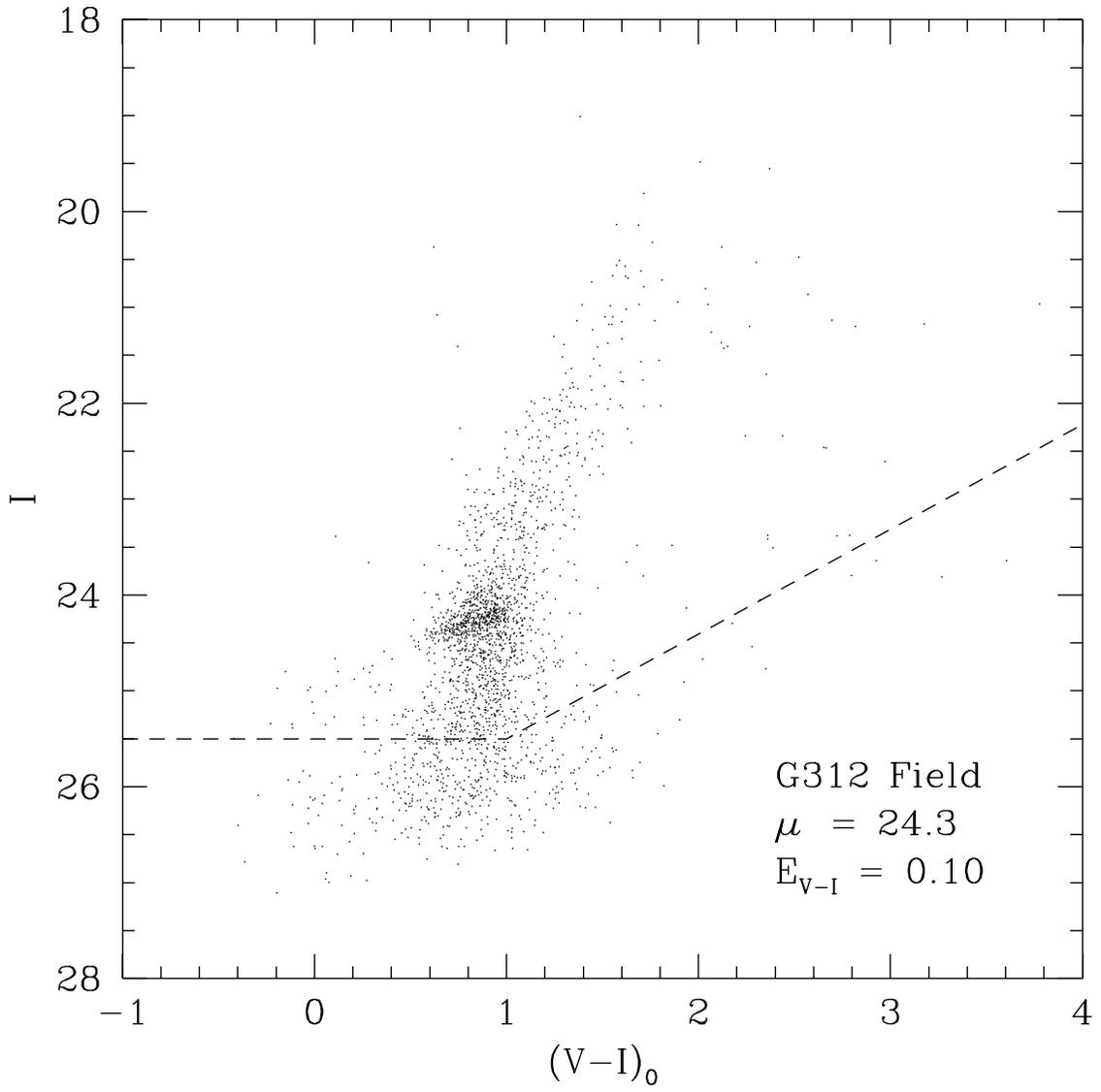}
\caption{This \fig is the same as \fig \ref{FIGURE:G302_cmd} but shows
the CMD for the M31 halo field around the GC G312. \label{FIGURE:G312_cmd}}
\end{figure}

	Figs. \ref{FIGURE:G302_cmd} and \ref{FIGURE:G312_cmd} show the
$(I,(V-I)_0)$ CMDs for the halo fields around the M31 GCs G302 and
G312 respectively.  No culling has been applied to these CMDs beyond
that done to identify real stellar images as described in
\sect{\ref{SECTION:data}}.  The CMD has been dereddened and corrected
for interstellar absorption.  The reader is refered to Holtzman \etal
\markcite{HB95}(1995b) for details on the corrections for interstellar
absorption.

	The CMDs of the M31 halo will contain contamination from
Galactic halo stars.  Galactic M-dwarfs have colors of $(V-I) \sim
+1.5$ to $+3$ so they will appear to the red of the RGB while
main-sequence stars and HB stars in the Galaxy will appear to the blue
of M31's RGB.  Star count models in the direction of M31, from
Ratnatunga \& Bahcall \markcite{RB85}(1985), suggest that there will
be $\sim 14 \pm 4$ stars redder than the RGB and $\sim 1 \pm 1$ stars
bluer than the RGB between $21 \le V \le 27$ in each of our fields.
Galactic field star contamination, therefore, contributes only
$\sim$0.2\% to the total number of stars observed in the G302 field
and $\sim$0.7\% to the G312 field.  The blue end of the HB is sparsely
populated but only $1 \pm 1$ foreground stars are expected on the blue
side of the RGB so uncertainties due to photometric scatter will
dominate over the effects of contamination.  The small number of
Galactic stars expected in our fields means that foreground
contamination is not a problem.

	Some faint background galaxies may have been mis-identified as
stars.  The deep galaxy counts of Smail \etal \markcite{SH95}(1995)
suggest that $19 \pm 2$ background galaxies with $20 \le I \le 24$ and
$168 \pm 20$ galaxies with $24 \le I \le 27$ should be located in each
of our two fields.  Typical galaxies have colors of $(V-I) \simeq 1.0$
so background galaxies are indistinguishable from stars in a CMD.  The
small expected number of galaxies with $I \le 24$ suggests that
background galaxies are not making a significant contribution to the
morphology of the RGB.  The blue end of the HB is $\sim$1 magnitude
bluer than typical background galaxies so we believe that the blue HB
is not significantly contaminated by galaxies.  The galaxy counts
suggest that for $I > 24$ background galaxies make up at most 2.9\%
$\pm$ 0.3\% of the objects on the G302 field CMD.  In the G312 field
background galaxies can account for up to 10.4\% $\pm$ 1.2\% of the
objects on the CMD.  {\sc DaoPhot/AllFrame} will discard objects that
are not morphologically similar to the stellar PSF for the frame in
question so many of the faint background galaxies will already have
been discarded.  Therefore we believe that mis-identified background
galaxies are not significantly biasing the distribution of stars on
the CMDs.

	A third potential source of contamination is the disk of M31
itself.  Mould \& Kristian \markcite{MK86}(1986) used the surface
photometry of de Vaucouleurs \markcite{dV58}(1958) to estimate that
the disk-to-halo ratio in a field $\sim$35\arcmin~from the center of
M31 along the SE minor axis was $\lesssim 0.014$.  Pritchet \& van den
Bergh \markcite{PvdB88}(1988) compared star counts in a field
40\arcmin~from the center of M31 along the SE minor axis with various
models and found that even with a thick disk component the
disk-to-halo ratio would be $\lesssim 0.04$.  Hodder
\markcite{Hod95}(1995) adapted the Bahcall \& Soneira \markcite{BS84}(1984)
Galaxy model to provide star counts and color distributions for an
external spiral galaxy and used this model to estimate a disk-to-halo
number-density ratio of $\sim$0.1 in the G302 field and $\sim$0.03 in
the G312 field, although these ratios are somewhat dependent on the
details of the disk and halo models.  Taking the worst-case scenario
for disk contamination (that of Hodder \markcite{Hod95}1995) leads us
to believe that less than 10\% of the stars in the G302 field, and
less than 3\% in the G312 field, are due to contamination from the
disk of M31.

	On the basis of these calculations we expect {\sl at most\/}
$\sim 13$\% of the objects in each field to be background galaxies or
stars that are not members of the M31 halo.  Therefore, we conclude
that the overall morphologies in our CMDs represent real features in
the halo population of M31.  The exception is the group of blue
objects visible $\sim$ 1 magnitude above the HB in the G302 field (see
\fig \ref{FIGURE:G302_cmd}).  A visual examination of the individual CCD
frames of the G302 field showed that approximately half of these
objects fall on or near background galaxies, saturated stars, or
pixels that has been flagged as bad for whatever reasons.  The
remaining $\sim$10 bright blue objects may be nucleated dwarf
galaxies, blends of stars, or stars with a large photometric
uncertainty.

\subsection{The Red Giant Branch\label{SECTION:rgb}}
	
	The RGBs in the G302 and G312 fields are morphologically
similar except for a clump of stars at $I \sim 23$ in the RGB of the
G302 field (see \fig \ref{FIGURE:G302_cmd}).  There is no apparent
corresponding clump visible in the G312 field (see \fig
\ref{FIGURE:G312_cmd}).  In order to check if this clump is a real
feature of the RGB we determined the differential LF for each RGB for
$22 \le V \le 24.5$ using a non-parametric histogram.  We then
constructed a second LF where the clump was removed by interpolating
the differential LF across the region containing the clump.  The
resulting ``clumpless'' LF was compared to the observed LF using a
two-sided two-sample Kolmogorov--Smirnov (K--S ) test.  This test said
that the density enhancement at $V \sim 24.25$ is significant at less
than the $5 \times 10^{-6}$\% confidence level, suggesting that the
clump in the RGB is not real.

	We applied the two-dimensional K--S test (Peacock
\markcite{P83}1983) to the entire RGB above $V \ge 24.5$ and found that
the RGB stars in the G302 and G312 fields were drawn from the same
distributions at the 91\% confidence level.  The same tests found that
the distributions of stars on the blue sides of the upper RGBs ($V \le
24$, $(V-I)_0 \le 1.6$) were the same at the 85\% confidence level and
that the red sides of the upper RGBs ($V \le 24$, $(V-I)_0 \ge 1.6$)
were the same at the 93\% confidence level.  This is weakly suggestive
that the two fields have the same stellar populations.

\begin{figure}
\plotone{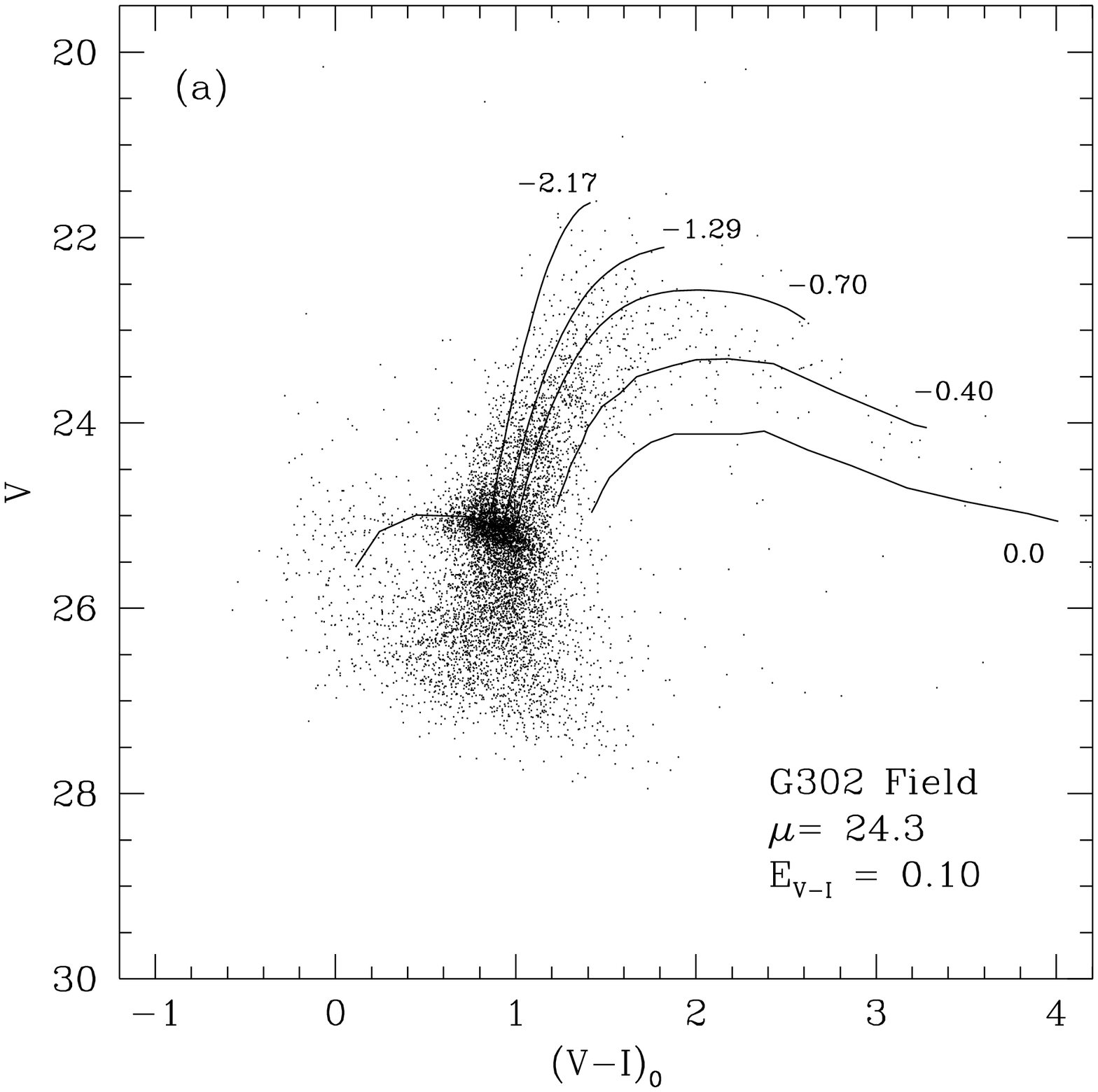}
\caption{This \fig shows the CMD for stars in the WF2 and the WF4 CCDs
in the G302 field.  The fiducial sequences are described in
\sect{\ref{SECTION:rgb}}. \label{FIGURE:G302_isochrones}}
\end{figure}

\begin{figure}
\plotone{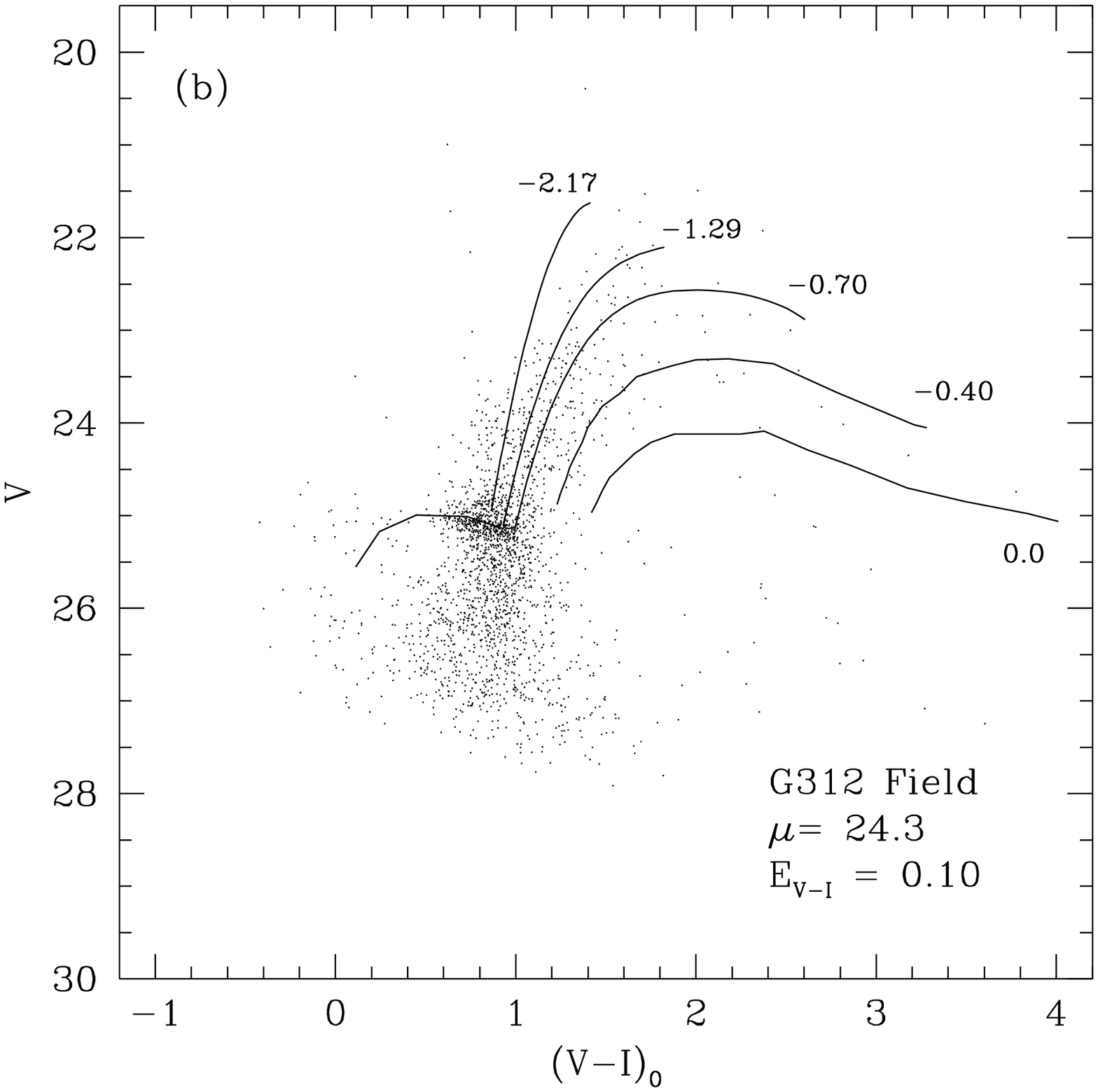}
\caption{This \fig shows the CMD for stars in the WF2 and the WF4 CCDs
in the G312 field.  The fiducial sequences are described in
\sect{\ref{SECTION:rgb}}. \label{FIGURE:G312_isochrones}}
\end{figure}

	\figs \ref{FIGURE:G302_isochrones} and
\ref{FIGURE:G312_isochrones} shows the $(V,(V-I)_0)$ CMDs for the M31
halo fields around G302 and G312 with a series of fiducial sequences
of different metallicities added.  The three metal-poor fiducial
sequences are the RGB ridge lines for the Galactic GCs M15 ($\mH =
-2.19$), NGC 1851 ($\mH = -1.29$), and 47 Tuc ($\mH = -0.71$) taken
from Da Costa \& Armandroff \markcite{DA90}(1990).  The two metal rich
fiducials are the $t_0 = 13.8$ Gyr isochrones of Bertelli \etal
\markcite{BB94}(1994) with $\mH = -0.4$ and $\mH = 0.0$.  The HB
fiducial sequence is that of M54 ($\mH = -1.42$) from Sarajedini \&
Layden \markcite{SL95}(1995).

	The observed distribution of stars along the red end of the
RGB is consistent with a metal-rich population while the blue side of
the RGB is consistent with a low-metallicity population.  However,
photometry of 47 Tuc (Lee \etal \markcite{L77}1977) shows that
$\sim$15\% of the stars $\sim$1 magnitude brighter than the HB are
evolved asymptotic giant branch (AGB) stars.  If this ratio holds for
the halo of M31 then many of the stars blueward of the $\mH = -1.29$
fiducial sequence will be AGB stars.  The spread in metallicity in the
RGB stars is clearly visible from the locations of the fiducial
sequences.  The majority of the stars, however, have $\rm{[m/H]} >
-0.7$, consistent with recent studies (see references in
\sect{\ref{SECTION:intro}}) which have found metallicities of
$\rm{[m/H]} \sim -0.6$ for the halo of M31.

	To estimate the contribution to the width of the RGB from the
finite depth of the halo we assumed that the projected surface density
distribution of stars in the M31 halo follows a de Vaucouleurs
$R^{1/4}$ law, $\Sigma(R)$, with an effective radius of $R_e = 1.3$
kpc and an axial ratio of $\alpha_s = 0.55$, in accordance with
Pritchet \& van den Bergh \markcite{PvdB94}(1994).  The volume
density, $\rho(r)$, can be obtained from $\Sigma(R)$, through Abel's
integral, as follows:

\begin{equation}
\rho(r) = -{1 \over \pi} \int_{R=r}^{+\infty} {d\Sigma(R) \over dR}
	       {dR \over \sqrt{R^2 - r^2}}
\end{equation}

\noindent
where $R$ is the projected distance of the field from the center of
M31 and $r$ is the true galactocentric distance of the volume element
$\rho(r)$.

	We computed the density distribution along the line-of-sight
at projected distances of 32\arcmin~and 50\arcmin~from the center of
M31.  The spread due to the depth of the halo has a half-power width
in \V~of $\sim$0.02 magnitudes at $R_{\rm M31} = 32 $\arcmin~(the G302
field) and $\sim$0.03 magnitudes at $R_{\rm M31} = 50$\arcmin~(the
G312 field).  Changing the effective radius and axial ratio over the
ranges found by Hodder \markcite{Hod95}(1995) in his study of the
structure of M31 does not significantly alter these results.  A change
of a few hundredths of a magnitude along any of the Da Costa \&
Armandroff \markcite{DA90}(1990) fiducial RGBs corresponds to a
negligible change in color so we conclude that the depth of the M31
halo does not contribute significantly to the observed width of the
RGB.

	The mean photometric uncertainties in our data are
$\sigma_{V-I} \simeq 0.05$ near the tip of the RGB ($I \simeq 20$) and
increase to $\sigma_{V-I} \simeq 0.10$ at the level of the HB yet the
observed spread in color along the RGB is $\sim$0.5 magnitudes at the
level of the HB and $\sim$2 magnitudes near $I \simeq 20$.  These
spreads are too large to be due to photometric uncertainties or the
depth of the M31 halo.  The turn-over in the RGB near $V = 23$ is due
to the increased opacity from molecular bands as giants expand and
become cooler.  For shell-hydrogen burning stars with $\mH \gtrsim -1$
the molecular opacity can become high enough to reduce the luminosity
of the star so that these stars become fainter as they ascend the RGB.
However, \figs \ref{FIGURE:G302_isochrones} and
\ref{FIGURE:G312_isochrones} shows that this flattening will not
explain all of the structure in the RGB.  Therefore a portion of the
width of the RGB is due to an intrinsic spread in the metallicity of
the M31 halo stars.

\begin{figure}
\plotone{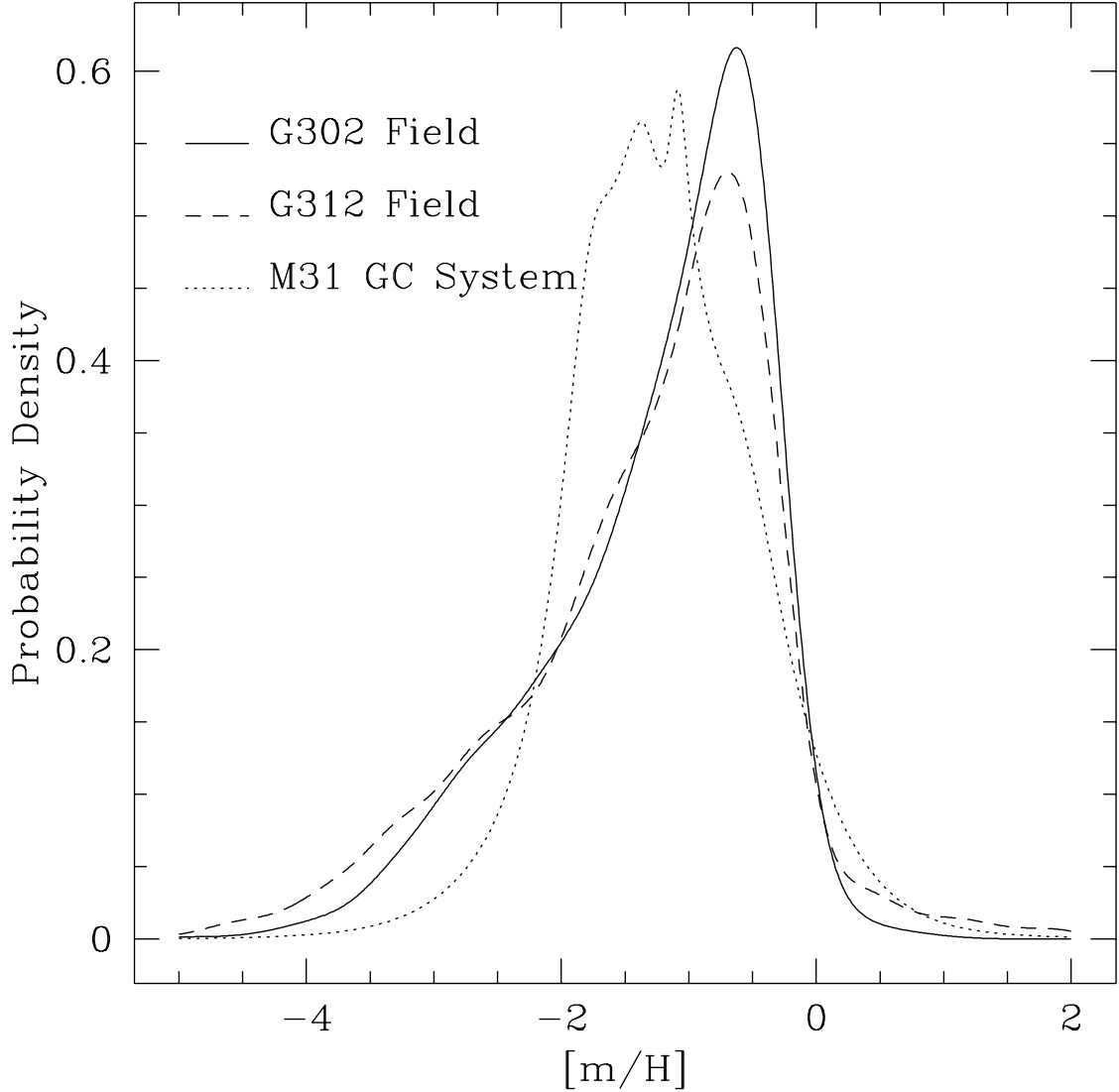}
\caption{The probability density distribution for the metallicity
of RGB stars in the G302 field (solid line), the G312 field (dashed
line), and for the M31 GC system (dotted line).  For each star we
generated a unit Gaussian with a standard deviation of $\sigma = 0.25$
dex, the estimated uncertainty in the metallicity determination for an
individual star.  For the GC distribution the individual $\sigma_\FeH$
values from Huchra \etal \protect\markcite{HB92}(1991) were used.  The
normalized sum of these Gaussians is a non-parametric histogram of the
probability density, $P(\mH)$, distribution.  The sharp red edge in
the distribution suggests that there is a well-defined upper limit to
the metallicity of stars in the G302 and G312 fields while the slow
decline on the blue edge suggests that the the metal-poor stars cover
a range of metallicities.  A portion of the metal-poor tail arises
from the confusion of AGB stars with metal-poor RGB
stars. \label{FIGURE:rgb_mdf}}
\end{figure}

	In order to estimate the metallicity distribution of RGB stars
we interpolated metallicity values for each star based on the fiducial
sequences in \figs \ref{FIGURE:G302_isochrones} and
\ref{FIGURE:G312_isochrones} and the theoretical isochrone of Bertelli
\etal \markcite{BB94}(1994) with $\mH = +0.4$ (not plotted).  We assumed
that the age of the M31 halo is $t_0 \simeq 14$ Gyr, comparable with
the age of the Galactic halo GC system (\eg Richer \etal
\markcite{RH96}1996).  The morphology of the RGB is not sensitive to
changes in age of a few Gyr so the exact age adopted for the RGB
isochrones is not critical.  \fig \ref{FIGURE:rgb_mdf} shows the
probability density distributions of metallicity ($\mH$) for RGB stars
with $I < 23$ in the G302 and G312 fields.  Based on the photometric
uncertainties in our data at the level of the RGB, and the
uncertainties inherent in matching theoretical isochrones to
observational data, we believe that our metallicity estimates for the
RGB stars have uncertainties of $\sigma_{\mH} \sim 0.25$.  A K--S test
shows that the two metallicity distributions differ at less than the
$5 \times 10^{-6}$\% confidence level, strongly suggesting that the
stellar populations are the same in both fields.  The G302 field
distribution has a peak at $\mH = -0.6$ and a FWHM of 1.3 dex while
the G312 field distribution has a peak at $\mH = -0.7$ and a FWHM of
1.6 dex.  These FWHMs are overestimates of the true metallicity
distribution since the probability density distribution function shown
in \fig \ref{FIGURE:rgb_mdf} is essentially a convolution of the data
with a unit Gaussian that has a dispersion equal to the uncertainty in
a typical measurement.  The intrinsic spread in the metallicity can be
estimated by deconvolving this Gaussian from the probability density
distribution function.  This gives an intrinsic FWHM of 1.2 dex for
the G302 field and 1.5 dex for the G312 field.  The halo stars are
clearly more metal-rich than, and have a slightly greater spread in
metallicity than, M31 GC system.

	The halo metallicity distribution is clearly asymmetric with
an extended metal-poor tail.  This is partly due to the presence of
AGB stars on the blue edge of the RGB and partly due to the
metallicities less than $\mH = -2.19$ being extrapolated from the
higher-metallicity fiducial sequences.  Therefore the shape of the
metallicity distribution beyond $\mH \sim -2$ should be regarded with
caution.  There is a sharp cut-off in the halo metallicity
distribution at approximately Solar metallicity. The metal-rich tail
of the metallicity distribution drops to zero quite rapidly between
$\mH \simeq -0.2$ and $\mH \simeq +0.2$. A visual examination of the
locations of the $\mH = -0.4$ and $\mH = 0.0$ isochrones in \figs
\ref{FIGURE:G302_isochrones} and \ref{FIGURE:G312_isochrones} shows
the most metal-rich RGB stars having $\mH \simeq -0.2$ suggesting that
the extra-Solar metallicity tail in \fig \ref{FIGURE:rgb_mdf} is an
artifact of the uncertainties in the metallicity determination.

\subsection{The Horizontal Branch\label{SECTION:hb}}

	The HB stars in both fields are concentrated in a red clump
between $0.5 \le (V-I)_0 \le 1.0$ with a handful of blue stars between
$0.0 \lesssim (V-I)_0 \le 0.5$.  The red clump is consistent with what
is seen in metal-rich Galactic GCs such as 47 Tuc while the small
number of blue HB stars is indicative of a metal-poor population.
However, an age spread of $\sim$ 5 Gyr in the halo population could
produce a HB morphology similar to what a metallicity spread of $\sim$
1 dex would produce.  This is the classic ``second parameter'' problem
(see van den Bergh \markcite{vdB93}1993, Lee \etal
\markcite{LD94}1994, and Chaboyer \etal \markcite{CD96}1996 for recent
reviews) affecting the colors of HB stars.  The lack of any extended
AGB candidates in our data, or in any other M31 halo fields, suggests
that a young population ( $\lesssim 5$ Gyr) is not present but there
is still the possibility that the M31 halo may contain stars as young
as $\sim$ 10 Gyr.

	The morphology of the HB in each field can be parameterized by
counting the number of red and blue HB stars and computing the
HB-ratio $ = (N_B - N_R) / (N_B + N_V + N_R)$ where $N_B$ is the
number of HB stars bluer than the instability strip, $N_V$ is the
number of stars in the instability strip, and $N_R$ is the number of
HB stars redder than the instability strip.  Since photometric scatter
is quite large ($\sigma_{(V-I)} \simeq 0.16$) at the level of the
instability strip, and the instability strip has a width of only
$\sim$0.25 magnitudes, we did not attempt to identify RR Lyrae
candidates.  Instead we split the HB at $(V-I)_0 = 0.5$ and calculated
the ratio $(N_B - N_R) / (N_B + N_R)$.  The G302 field has a HB-ratio
of $-0.91 \pm 0.12$ (Poisson uncertainty) while the G312 field has a
HB-ratio of $-0.92 \pm 0.26$ suggesting that both fields have the same
HB morphology.  These ratios are lower limits on the actual HR-ratios
since the blue HB extends to below the $\sigma_{(V-I)} = 0.2$
photometric limit suggesting that incompleteness may be causing us to
seriously underestimate $N_B$.  We used the theoretical HBs of Lee
\etal \markcite{LD94}(1994) to estimate the expected values of the HB
type given the metallicity ratios determined in
\sect{\ref{SECTION:lf}}.  The predicted HB-ratios were $-0.6 \pm 0.3$
in the G302 field and $-0.1 \pm 0.1$ in the G312 field, somewhat more
positive than the values measured from the CMDs.  This suggests that
either the blue HB suffers from a high degree of incompleteness, or
that there is a second parameter contributing to the morphology of the
HB in the halo of M31.  Detailed incompleteness tests are in progress
to attempt to determine whether or not our results are due to the
second parameter effect.  At present we conclude that the morphology
of the HB in the halo of M31 is broad agreement with the metallicity
distribution determined from the RGB morphology but that further work
is needed to determine if the lack of blue HB stars (relative to the
metallicity spread observed in the RGB stars) is due to photometric
limits in our data or a second parameter effect.

	The CMDs (\figs \ref{FIGURE:G302_isochrones} and
\ref{FIGURE:G312_isochrones}) shows that the HB is not actually horizontal in
$V$.  To estimate the spread in HB magnitudes we split the HB into
bins with widths of $\Delta (V-I) = 0.1$ and fit a sloping Gaussian to
the distribution of stars in $V$ in each bin.  The mean width of these
Gaussians was $\sigma_V = 0.14 \pm 0.01$ (standard error) in the G302
field and $\sigma_V = 0.12 \pm 0.03$ in the G312 field.  The mean
photometric uncertainty at the level of the red HB is $\sigma_V =
0.086$.  Subtracting this quadratically from the observed spread in
\V-magnitude of the HB gives intrinsic spreads of $\sigma_V = 0.11 \pm
0.01$ in the G302 field and $\sigma_V = 0.08 \pm 0.03$ in the G312
field.  The finite depth of the halo of M31 also introduces a small
broadening in \V~(see \sect{\ref{SECTION:rgb}}) of the HB.  However,
this broadening is negligible ($\sim$3\%) compared with the broadening
introduced by photometric uncertainties so it was ignored.  The
resulting distribution of \V-band magnitudes was converted to iron
abundances using the relation of Chaboyer \etal \markcite{CD96}(1996),
see below.  The derived mean iron abundance is $\FeH = -0.5 \pm 0.6$
(standard deviation) in the G302 field and $\FeH = -0.5
\pm 0.4$ in the G312 field.

	The mean magnitude of the red HB stars is $\overline{V} =
25.18 \pm 0.01$ in the G302 field and $\overline{V} = 25.17 \pm 0.01$
in the G312 field.  These correspond to absolute magnitudes for the HB
of $M_V(HB) = 0.88 \pm 0.1$ in the G302 field and $M_V(HB) = 0.87 \pm
0.1$ in the G312 field.  The relationship between metallicity and
absolute magnitudes of RR Lyrae variables, $M_V(RR)$, is somewhat
uncertain.  Chaboyer \etal \markcite{CD96}(1996) find $M_V(RR) =
0.20\FeH + 0.98$, while Carney \etal \markcite{CS92}(1992) find
$\langle M_V(RR) \rangle = (0.15 \pm 0.01)\FeH + (1.01 \pm 0.08)$.
The spread in \mH~on the RGB, and the observed spread in the magnitude
of the red HB stars, allows us to make an estimate of the slope of the
metallicity--magnitude relation for HB stars.  If we assume that the
slope ($d M_V / d \mH$) is positive and combine the data from both
fields to improve the statistics we derive a relationship of $M_V(HB)
= (0.22 \pm 0.02)\mH + (1.01 \pm 0.11)$.  A preliminary analysis of
the CMDs of the M31 GCs G302 and G312 suggests $M_V(HB) = 0.2\mH +
1.0$ (see Holland \etal\markcite{HF96} 1996 for details).  These
results are consistent with the Chaboyer \etal \markcite{CD96}(1996)
relation.

	The relationship between iron abundance and HB magnitude is
very sensitive the the adopted value of the distance modulus.  An
error of only 0.02 magnitudes in the distance modulus will result in
an uncertainty of $\sim 0.1$ dex in metallicity.  Therefore, we
believe that the metallicity distribution estimated from the RGB is a
better indicator of the mean metallicity of the M31 halo than the
metallicity distribution obtained from the HB.  However, it should be
noted that the metal-poor edge of the HB metallicity distribution
($\overline{\FeH} - 3\sigma_\FeH$) occurs at $\FeH \simeq -2$ in rough
agreement with the metallicity distribution of RGB stars.  The lack of
HB stars with $\FeH \lesssim -2$ supports our conclusion that the
metal-poor tail of the RGB metallicity distribution is actually due to
calibration uncertainties and contamination of the blue side of the
RGB by AGB stars.

	The metal-rich edge of the HB metallicity distribution
($\overline{\FeH} + 3\sigma_\FeH$) occurs at $\FeH \simeq +1$.  The
RGBs show no evidence for stars with $\FeH \gtrsim -0.2$ so the
metal-rich tail in the HB metallicity distribution is most likely an
artifact of fitting a Gaussian to a non-Gaussian distribution.  In
general, red HBs have a well defined faint edge but photometric
uncertainties and contamination from RGB stars make it impossible to
identify this faint edge in our data.  Therefore we conclude that the
metal-rich tail of the HB metallicity distribution is spurious.

\subsection{The Helium Abundance\label{SECTION:helium}}

	The helium abundance in the halo of M31 can be estimated from
the relative numbers of stars on the HB, the RGB and the AGB,
$R^\prime = N_{\rm HB} / (N_{\rm RGB} + N_{\rm AGB})$ (\eg Buzzoni
\etal \markcite{BF83}1983).  The number of RGB stars brighter than the HB
was counted and an estimate of the number of Galactic foreground stars
and background galaxies (see \sect{\ref{SECTION:contam}}) was
subtracted.  The number of HB stars was difficult to estimates since:
(1) the blue end of the HB is near the photometric limit of the data,
and (2) the red clump of the HB sits on top of the RGB.  Since $N_B$
is small compared with $N_R$ ($N_R \sim 0.95 N_B$), and $N_B$ remains
approximately constant with color bluewards of $(V-I)_0 \simeq 0.5$,
we believe that the uncertainty in the number of blue HB stars will
not significantly affect the determination of $N_{\rm HB}$.  To
estimate the degree of contamination in the red clump due to the RGB
we interpolated the RGB LF across the HB and subtracted these stars
from $N_{\rm HB}$.  Since there is confusion between AGB and RGB stars
we used the $R^\prime$ estimator of helium abundance instead of the $R
= N_{\rm HB} / N_{\rm RGB}$ estimator.

	The resulting ratios are $R^\prime = 1.37 \pm 0.08$ in the
G302 field and $R^\prime = 0.98 \pm 0.11$ in the G312 field.  Using
the calibration of Buzzoni \etal \markcite{BF83}(1983) these ratios
yield helium abundances of $Y = 0.27 \pm 0.03$ in the G302 field and
$Y = 0.20 \pm 0.05$ in the G312 field.  Our uncertainty estimates only
include Poisson uncertainties and an estimate of the uncertainty in
our estimate of the RGB contamination in the red clump of the HB.
They do not include any potential incompleteness effects at the blue
end of the HB.  Therefore, our quoted uncertainties should be treated
as lower limits on the true uncertainty.  Further, the effects of
incompleteness as the blue end of the HB may have resulted in our
underestimating the number of HB stars and therefore underestimating
the helium abundance.  Since the G302 field contains $\sim$4 times the
number of stars as the G312 field we believe that the helium abundance
derived from the G302 field CMD is a more reliable estimate of the
helium abundance in the halo of M31 than that derived from the G312
field CMD.  We note that the two values are in agreement at the $\sim$
1-$\sigma$ level.

	A halo helium abundance of $Y \simeq 0.20$ to 0.27 is
consistent with the mean helium abundance, $\overline{Y} = 0.23 \pm
0.02$ (Buzzoni \etal\markcite{BF83} 1983), seen in the blue Galactic
globular clusters.  Our estimate is slightly less than the helium
abundance of 47 Tuc, $Y = 0.28 \pm 0.04$ (Hesser \etal
\markcite{HH87}1987), however the large uncertainty in our result
makes it compatible with that of 47 Tuc.  The similarity between the
helium abundance of the old Galactic GCs and the halo stars in M31
suggests that the two populations are similar.

%%%%%%%%%%%%%%%%%%%%%%%%%%%%%%%%%%%%%%%%%%%%%%%%%%%%%%%%%%%%%%%%%%%%%

\section{The Halo Luminosity Function\label{SECTION:lf}}

\begin{figure}
\plotone{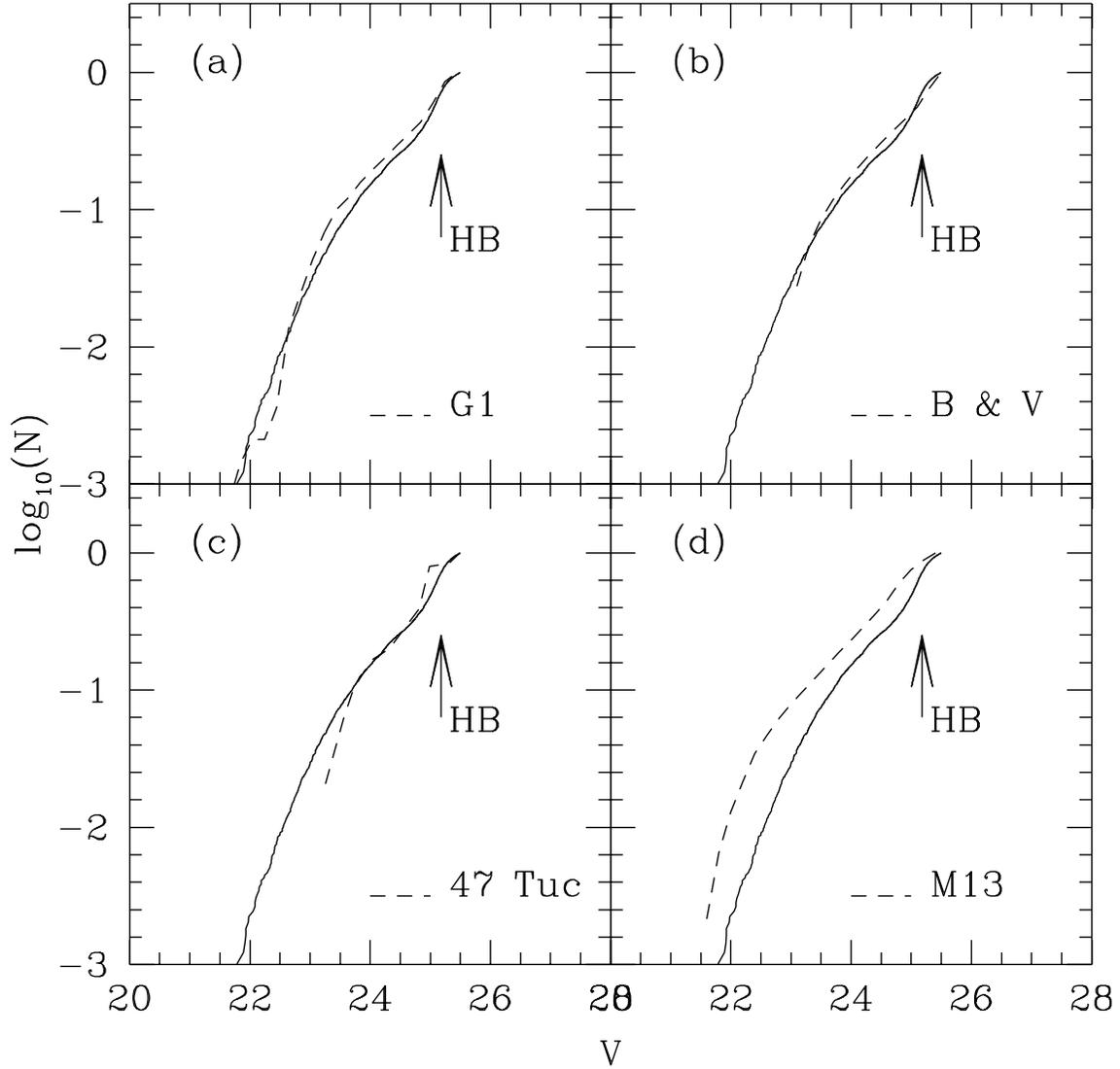}
\caption{Here we compare the cumulative LF for the M31 halo field
near G302 (solid lines) with four comparison LFs (dashed lines: (a)
the G1 LF, (b) a theoretical LF (B \& V) with $\FeH = -0.47$, $\OFe =
+0.23$ and $t_0 = 14$ Gyr, (c) the 47 Tuc LF, and (d) the M13 LF.  The
comparison LFs were scaled so that the total number of stars with $21
\le V \le 25.5$ was the same as the number of stars in the G302 field
in the same range of magnitudes.  $V = 25.5$ was selected as the faint
cut-off because we believe our photometry is reasonably complete down
to approximately this point
(see \sect{\ref{SECTION:lf}}). \label{FIGURE:G302_lfs}}
\end{figure}

\begin{figure}
\plotone{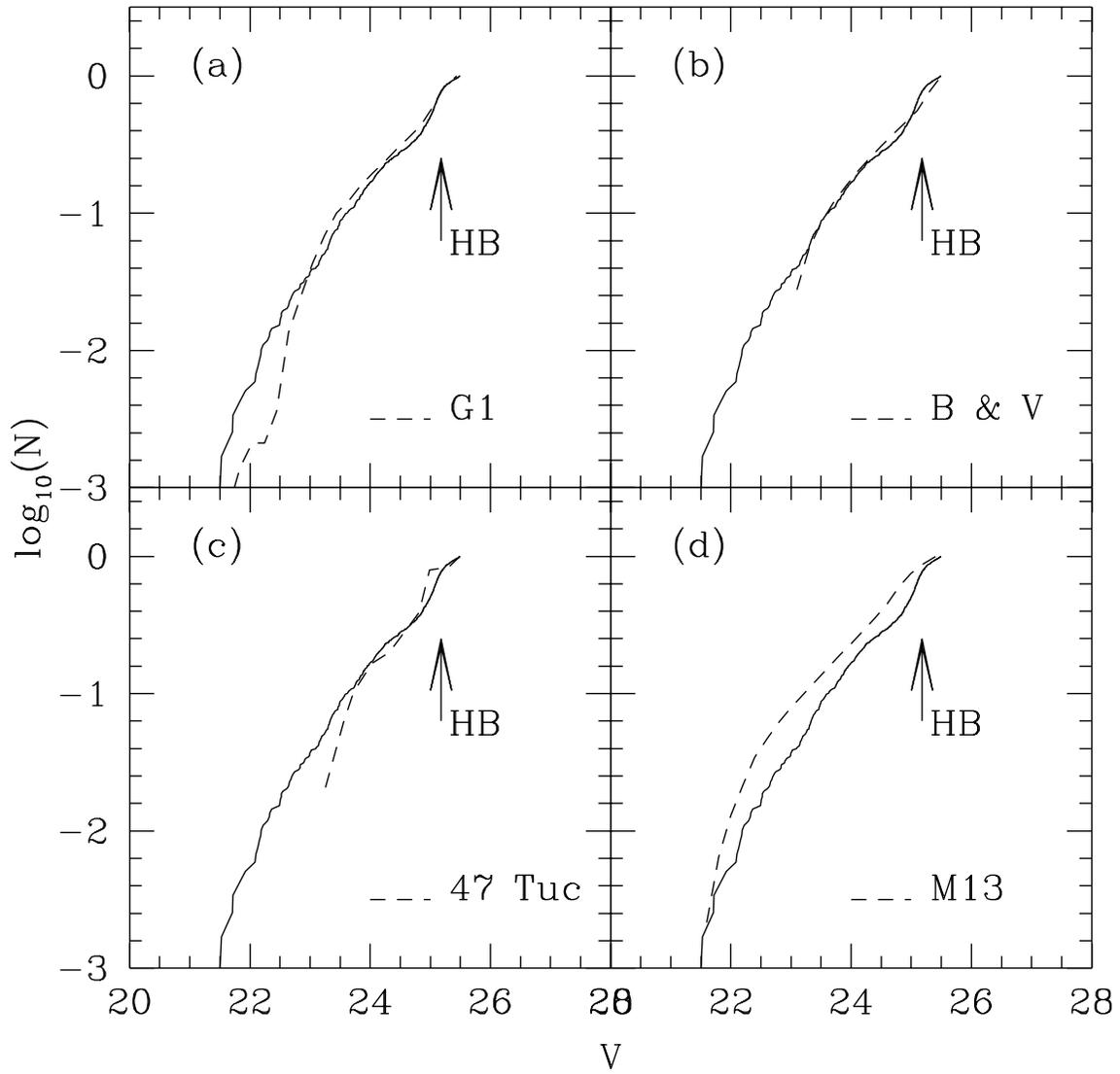}
\caption{Here we compare the cumulative LF for the M31 halo field
near G312 with the four comparison LFs used in \fig
\ref{FIGURE:G302_lfs}. \label{FIGURE:G312_lfs}}
\end{figure}

	Preliminary artificial star tests indicate that our photometry
is complete to at least the level of the HB ($V \simeq 25$) in both
the G302 and G312 fields.  This is consistent with the results of Rich
\etal \markcite{RM96}(1996) who found that star-counts in the halo of M31
obtained using WFPC2 data were complete to at least $V \sim 26$.  The
\V- and \I-band LFs in each field are tabulated in Table \ref{TABLE:lum_fun}.
The $\phi$ values are the number of stars in each field per 0.2
magnitude bin.  The total area of each field is $\sim 3.2 \sq\arcmin$
after masking and bad pixels are taken into account.  \fig
\ref{FIGURE:G302_lfs} shows the LF for the G302 field compared with
LFs for the metal-rich M31 GC G1 (Rich \etal \markcite{RM96}1996), a
theoretical LF for an old ($t_0 = 14$ Gyr) metal-rich ($\FeH = -0.47$,
$\OFe = +0.23$) population (Bergbusch \& VandenBerg
\markcite{BV92}1992), the metal-rich Galactic GC 47 Tuc (Hesser
\etal \markcite{HH87}1987), and the metal-poor Galactic GC M13 (Simoda
\& Kimura \markcite{SK68}1968).  The comparison LFs were scaled so
that they had the same number of stars with $21.4 \le V \le 25.5$ as
the observed M31 halo LF did.  In all four cases K--S tests returned
less than a 17\% confidence that the G302 field LF is drawn from one
of the comparison single-metallicity populations.  \fig
\ref{FIGURE:G312_lfs} show the LF for the G312 field and the same
comparison LFs.  K--S tests returned less than a 17\% confidence that
the G312 LF is drawn from the same population as any of the comparison
LFs.

\begin{figure}
\plotone{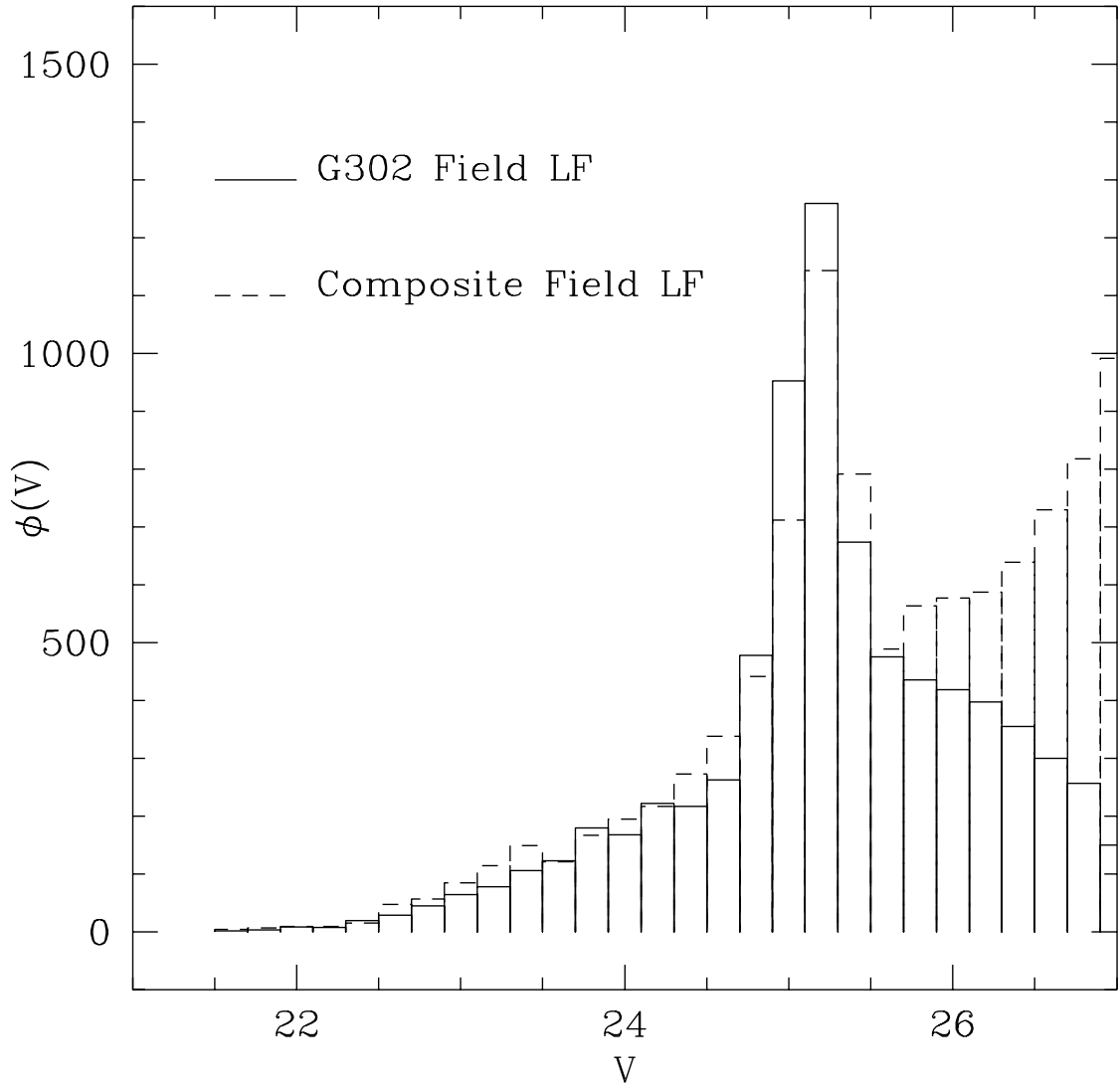}
\caption{The observed LF in the G302 field and the composite LF
built from scaling the G1 and M13 LFs as described in
\sect{\ref{SECTION:lf}}.  The poor agreement at $V \gtrsim 25.5$ may
be due to incompleteness in the observed LF. \label{FIGURE:G302_lf_mixed}}
\end{figure}

\begin{figure}
\plotone{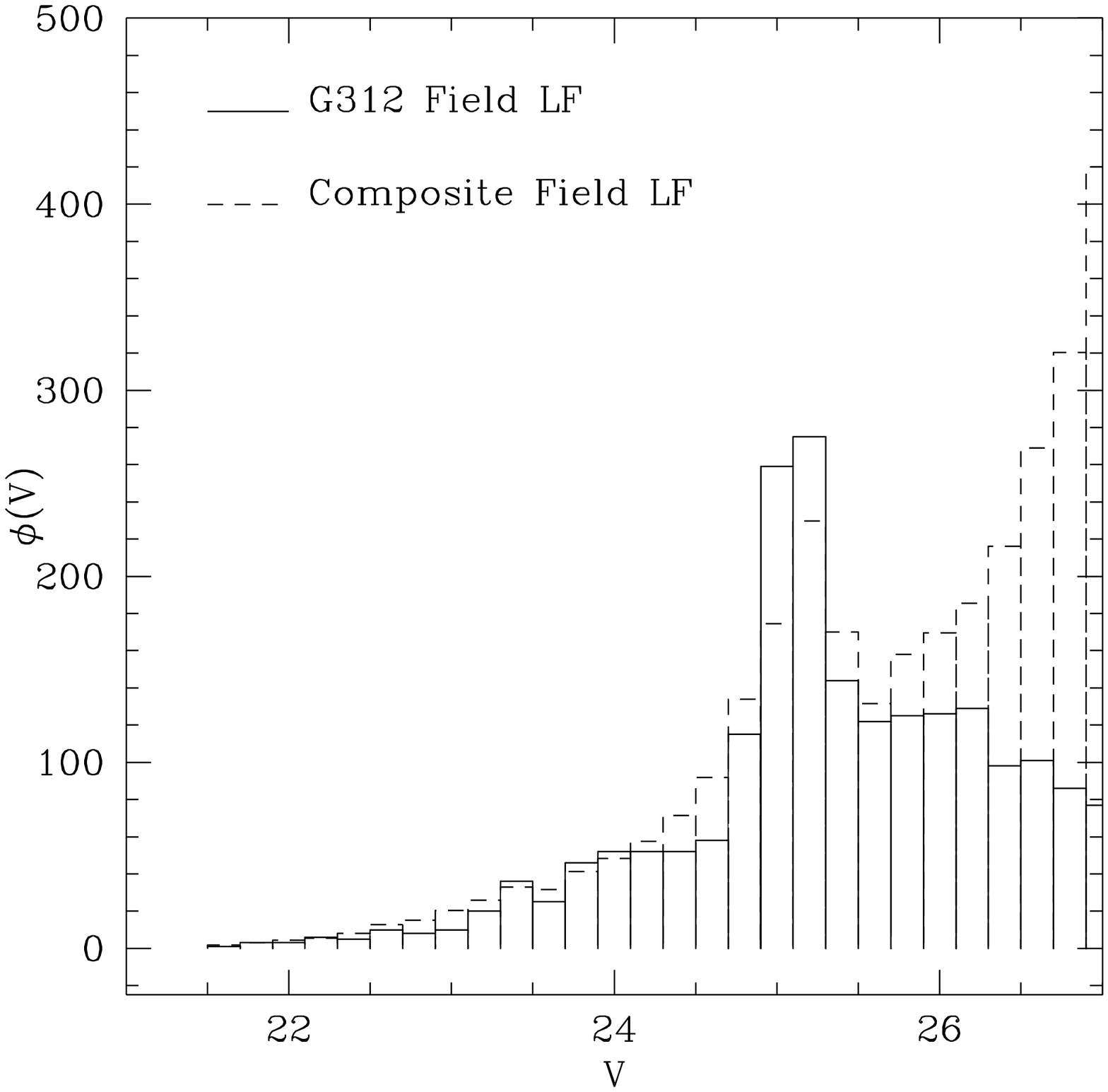}
\caption{The same as \fig \ref{FIGURE:G302_lf_mixed}
but for the G312 field LF. \label{FIGURE:G312_lf_mixed}}
\end{figure}

	It would be of interest to determine what fraction of the M31
halo stars belong to the metal-rich ($\mH \sim -0.6$) population and
what fraction belong to the metal-poor popu lation that gives rise to
the RR Lyrae stars.  In order to estimate this ratio we modeled the
observed LF of the M31 halo using a composite LF built from a
metal-poor LF and a metal-rich LF.  To model the metal-poor population
we took the LF of the Galactic GC M13 ($=$ NGC 6205), a typical
metal-poor GC with $\FeH = -1.65$ (Zinn \& West \markcite{ZW84}1984).
For the metal-rich population we used the observed LF of the M31 GC G1
from Rich \etal \markcite{RM96}(1996).  G1's CMD suggests that cluster
has a metallicity of $\FeH \simeq -0.6$ (Rich \etal
\markcite{RM96}1996), significantly greater than the spectroscopic
metallicity (Huchra \etal \markcite{HB91}1991) of $\FeH = -1.09$.  To
determine the ratio of metal-poor to metal-rich stars we assumed that
all the observed M31 halo stars stars with $V < 22$ belonged to the
metal-poor population.  An examination of the CMDs (\figs
\ref{FIGURE:G302_isochrones} and \ref{FIGURE:G312_isochrones}) shows
no evidence for a significant number of RGB stars with metallicities
less than this.  We normalized the M13 LF to this number then scaled
the G1 LF so that the total number of metal-poor and metal-rich stars
(M13 LF $+$ G1 LF) equaled the observed number of M31 halo stars with
magnitudes of $21.5 \le V \le 25.5$.  \fig \ref{FIGURE:G302_lf_mixed}
shows the observed differential LF for the G302 field and the
differential composite LF. \fig \ref{FIGURE:G312_lf_mixed} shows the
same thing for the G312 field LF.

	The observed number of stars drops relative to the composite
LF when $V \gtrsim 25.5$ suggesting that incompleteness is becoming
significant at these magnitudes.  K--S tests indicate that the
composite LFs are the same as the observed LFs at the 99.97\%
confidence level for $V \le 25$ suggesting that the ratio of
metal-rich to metal-poor populations is, broadly speaking, in
agreement with the actual distribution in each field.  The LFs were
not compared at $V > 25$ since there is the possibility that a second
parameter affecting the morphology of the HB.  In the G302 field the
ratio of the metal-rich LF to the metal-poor LF is $3.58 : 1$ while in
the G312 field this ratio is $0.95 : 1$.  While it is tempting to
conclude that that the G312 field contains a significantly different
mix of stellar populations than the G302 field does it must be
remembered that the metal-poor LF in our composite LF was scaled based
on the (small) number of stars with $21.5 \le V \le 22.5$.  Of the
$\sim$10 stars observed in this magnitude range in the G302 field
$\sim$ 2 will be foreground stars (Ratnatunga \& Bahcall
\markcite{RB85}1985) and some will be AGB stars.  These effects, and
Poisson uncertainties, will introduce uncertainties of $\sim$50\% into
the derived metallicity ratios.

	The large uncertainties in our composite LFs make us hesitant
to make any definite statements as to whether or not the stellar
populations in our two fields are the same.  The composite LFs in the
two fields are the same at only the 9.8\% confidence level, however
the RGB and HB morphologies and metallicity distributions all suggest
that the G302 field and G312 field contain the same stellar
populations.  We do not, therefore, believe that our analysis
conclusively demonstrates that the G302 field contains a different mix
of stellar populations than the G312 field does but we stress that
this is not a strong conclusion and that further work will be needed
in many different halo fields to determine what, if any, variation
between position and population exists in the halo of M31.  We do,
however, believe that this analysis demonstrates that a non-negligible
fraction ($\sim$ 25\% to 50\%) of the halo of M31 {\sl in our fields
of study\/} is made up of metal-poor stars that may give rise to the
observed RR Lyrae variables.

%%%%%%%%%%%%%%%%%%%%%%%%%%%%%%%%%%%%%%%%%%%%%%%%%%%%%%%%%%%%%%%%%%%%%%%%%%%%%%

\section{Conclusions\label{SECTION:conc}}

	We used {\sl HST\/} WFPC2 \V- and \I-band photometry to show
that the stellar halo of M31 has a mean metallicity of $\mH \simeq
-0.6$ with a spread of $-2 \lesssim \mH \lesssim -0.2$, comparable to
the metallicity determinations made using ground-based observations.
This result is primarily based on a comparison of the morphology of
the RGB to fiducial sequences of Galactic GCs and theoretical RGB
isochrones.  The HB consists primarily of red clump stars, similar to
what is seen in 47 Tuc, and a small number of blue stars.  A
metallicity of $\mH \simeq -0.6$, the observed metallicity spread in
the RGB, and the observed distribution of stellar luminosities on the
HB is consistent with the Chaboyer \etal \markcite{CD96}(1996)
relationships between the metallicity and absolute magnitudes of RR
Lyrae variables.  The horizontal branch morphology ratio is $(N_B -
N_R) / (N_B + N_R) \simeq -0.9$, suggesting that there is a metal-poor
population present that could give rise to the RR Lyrae variables seen
by Pritchet \& van den Bergh \markcite{PvdB87}(1987).  The derived
helium abundance is $Y \simeq 0.20$ to $0.27$, similar to that in
Galactic GCs.

	Combining LFs from metal-rich and metal-poor populations and
comparing the resulting composite LF to the observed LF for the M31
halo suggests that metal-rich ($\mH \simeq -0.6$) stars make up
between half and three-quarters of the M31 halo in each of our fields
while the remaining stars belong to a metal-poor population with $\mH
\sim -1.5$ to $-2.0$.  The metallicity distributions of HB and RGB
stars are both consistent with such a mix of metal-rich and metal-poor
stars.  The HB, however, is redder than would be expected given this
mix of metallicities.  This may be due to blue HB stars being too
faint to appear in our data or it may indicate that a second parameter
is affecting the HB morphology in the halo of M31.

	We find no conclusive evidence for a difference in stellar
population between the G302 field and the G312 field.  LFs suggest
that the G312 ($R_{\rm M31} = 50$\arcmin) field may contain a higher
fraction of metal-poor stars than the G302 ($R_{\rm M31} = 32$\arcmin)
field but the metallicity distribution of RGB stars indicates that
there is no significant difference in the ratio of metal-rich to
metal-poor stars in the two fields.  Deep photometry will be needed in
a large number of fields in the M31 halo to determine how (or if)
metallicity varies with position in the halo of M31.

%%%%%%%%%%%%%%%%%%%%%%%%%%%%%%%%%%%%%%%%%%%%%%%%%%%%%%%%%%%%%%%%%%%%%%%%%%%%%%

\acknowledgments

	This research is based on observations made with the NASA/ESA
{\sl Hubble Space Telescope\/} obtained at the Space Telescope Science
Institute.  STScI is operated by the Association of Universities for
Research in Astronomy Inc.\ under NASA contract NAS 5-26555.  Support
for this research was provided by operating grants to GGF and HBR from
the Natural Science and Engineering Research Council of Canada.  SH
would like to thank Peter Stetson for kindly making a copy of the {\sc
AllFrame} software available.  We would also like to thank the
referee, Ren{\'e} Racine, for making several useful suggestion.

%%%%%%%%%%%%%%%%%%%%%%%%%%%%%%%%%%%%%%%%%%%%%%%%%%%%%%%%%%%%%%%%%%%%%%%%%%%%%%

%%%%%%%%%%%%%%%%%%%%%%%%%%%%%%%%%%%%%%%%%%%%%%%%%%%%%%%%%%%%%%%%%%%%%%%%%%%

\newpage

\begin{deluxetable}{rlcrcl}
%\tablewidth{181.55586pt}
\tablewidth{0pt}
\tablenum{1}
\tablecaption{Log of the Observations.\label{TABLE:obs_log}}
\tablehead
{	\colhead{Field} &
	\colhead{Date} &
	\colhead{Filter} &
	\colhead{Exposure} \nl
	\colhead{} &
	\colhead{(1995)} &
	\colhead{ } &
	\colhead{(s)}
}
\startdata
G302 & Nov.\ 5  & F555W  & 8 $\times$ 500  \nl
     &          &        & 2 $\times$ 160  \nl
     &          & F814W  & 7 $\times$ 500  \nl
     &          &        & 1 $\times$ 400  \nl
     &          &        & 1 $\times$ 160  \nl
G312 & Oct.\ 31 & F555W  & 8 $\times$ 500  \nl
     &          &        & 2 $\times$ 160  \nl
     &          & F814W  & 7 $\times$ 500  \nl
     &          &        & 1 $\times$ 400  \nl
     &          &        & 1 $\times$ 160  \nl
\enddata
\end{deluxetable}

%%%%%%%%%%%%%%%%%%%%%%%%%%%%%%%%%%%%%%%%%%%%%%%%%%%%%%%%%%%%%%%%%%%%%%%%%%%

\begin{deluxetable}{rrrccr}
%\tablewidth{272.13927pt}
\tablewidth{0pt}
\tablenum{2}
\tablecaption{Aperture Corrections.\label{TABLE:apcor}}
\tablehead
{	\colhead{Field} &
	\colhead{CCD} &
	\colhead{Filter} &
	\colhead{$<{\rm ap} - {\rm PSF}>$} &
	\colhead{Slope} &
	\colhead{N}
}
\startdata
G302 & WF2 & F555W & $+0.0184 \pm 0.0216$ & 1.065 &  25 \nl
     &     & F814W & $+0.0374 \pm 0.0099$ & 1.017 &  80 \nl
     & WF4 & F555W & $+0.0333 \pm 0.0119$ & 1.011 &  90 \nl
     &     & F814W & $+0.0297 \pm 0.0053$ & 1.005 & 252 \nl
G312 & WF2 & F555W & $-0.0344 \pm 0.0144$ & 1.008 & 186 \nl
     &     & F814W & $+0.0405 \pm 0.0087$ & 1.034 & 119 \nl
     & WF4 & F555W & $+0.0513 \pm 0.0165$ & 1.001 &  85 \nl
     &     & F814W & $+0.0333 \pm 0.0102$ & 1.027 &  89 \nl
\enddata
\end{deluxetable}

%%%%%%%%%%%%%%%%%%%%%%%%%%%%%%%%%%%%%%%%%%%%%%%%%%%%%%%%%%%%%%%%%%%%%%%%%%%

\begin{deluxetable}{ccccccc}
%\tablewidth{248.38931pt}
\tablewidth{0pt}
\tablenum{3}
\tablecaption{Photometric Uncertainties.\label{TABLE:errs}}
\tablehead
{	\colhead{} &
	\colhead{G302F} &
	\colhead{G312F} &
	\colhead{} &
	\colhead{G302F} &
	\colhead{G312F} \nl
	\colhead{$V$} &
	\colhead{$\sigma_V$} &
	\colhead{$\sigma_V$} &
	\colhead{$I$} &
	\colhead{$\sigma_I$} &
	\colhead{$\sigma_I$}
}
\startdata
 20.25 &    0.035 &  0.033 &    20.25 &    0.033 &    0.028 \nl
 20.75 &    0.033 &  0.038 &    20.75 &    0.028 &    0.029 \nl
 21.25 &  \nodata &  0.031 &    21.25 &    0.030 &    0.028 \nl
 21.75 &    0.050 &  0.035 &    21.75 &    0.031 &    0.031 \nl
 22.25 &    0.043 &  0.038 &    22.25 &    0.034 &    0.034 \nl
 22.75 &    0.039 &  0.042 &    22.75 &    0.041 &    0.040 \nl
 23.25 &    0.045 &  0.045 &    23.25 &    0.049 &    0.048 \nl
 23.75 &    0.053 &  0.053 &    23.75 &    0.061 &    0.060 \nl
 24.25 &    0.062 &  0.061 &    24.25 &    0.075 &    0.073 \nl
 24.75 &    0.076 &  0.076 &    24.75 &    0.100 &    0.100 \nl
 25.25 &    0.091 &  0.090 &    25.25 &    0.142 &    0.144 \nl
 25.75 &    0.121 &  0.122 &    25.75 &    0.208 &    0.205 \nl
 26.25 &    0.165 &  0.166 &    26.25 &    0.299 &    0.304 \nl
 26.75 &    0.235 &  0.241 &    26.75 &    0.463 &    0.459 \nl
 27.25 &    0.344 &  0.346 &    27.25 &    0.752 &    0.677 \nl
 27.75 &    0.547 &  0.492 &    27.75 &  \nodata &  \nodata \nl
\enddata
\end{deluxetable}

%%%%%%%%%%%%%%%%%%%%%%%%%%%%%%%%%%%%%%%%%%%%%%%%%%%%%%%%%%%%%%%%%%%%%%%%%%

\begin{deluxetable}{rrrrrr}
%\tablewidth{248.38931pt}
\tablewidth{0pt}
\tablenum{4}
\tablecaption{The {\sl V\/}- and {\sl I\/}-band apparent
               LFs.\label{TABLE:lum_fun}}
\tablehead
{	\colhead{} &
	\colhead{$\phi_V$} &
	\colhead{$\phi_V$} &
	\colhead{} &
	\colhead{$\phi_I$} &
	\colhead{$\phi_I$} \nl
	\colhead{$V$} &
	\colhead{G302F} &
	\colhead{G312F} &
	\colhead{$I$} &
	\colhead{G302F} &
	\colhead{G312F}
}
\startdata
 20.00 &    0 &     0 &   20.00 &    2 &     0  \nl
 20.20 &    2 &     0 &   20.20 &    4 &     2  \nl
 20.40 &    1 &     1 &   20.40 &    8 &     4  \nl
 20.60 &    1 &     0 &   20.60 &   17 &     9  \nl
 20.80 &    0 &     0 &   20.80 &   35 &     5  \nl
 21.00 &    1 &     1 &   21.00 &   42 &     9  \nl
 21.20 &    0 &     0 &   21.20 &   53 &    12  \nl
 21.40 &    0 &     1 &   21.40 &   49 &     9  \nl
 21.60 &    2 &     1 &   21.60 &   46 &     9  \nl
 21.80 &    4 &     3 &   21.80 &   49 &    11  \nl
 22.00 &    9 &     3 &   22.00 &   71 &    25  \nl
 22.20 &    8 &     6 &   22.20 &   69 &    18  \nl
 22.40 &   20 &     5 &   22.40 &  113 &    28  \nl
 22.60 &   29 &    10 &   22.60 &   89 &    28  \nl
 22.80 &   45 &     8 &   22.80 &  121 &    30  \nl
 23.00 &   65 &    10 &   23.00 &  154 &    42  \nl
 23.20 &   78 &    20 &   23.20 &  194 &    52  \nl
 23.40 &  106 &    36 &   23.40 &  174 &    45  \nl
 23.60 &  123 &    25 &   23.60 &  225 &    44  \nl
 23.80 &  180 &    46 &   23.80 &  304 &    68  \nl
 24.00 &  168 &    52 &   24.00 &  555 &   138  \nl
 24.20 &  222 &    52 &   24.20 & 1204 &   275  \nl
 24.40 &  217 &    52 &   24.40 &  987 &   237  \nl
 24.60 &  263 &    58 &   24.60 &  586 &   145  \nl
 24.80 &  478 &   115 &   24.80 &  418 &   106  \nl
 25.00 &  953 &   259 &   25.00 &  388 &   110  \nl
 25.20 & 1260 &   275 &   25.20 &  357 &   115  \nl
 25.40 &  674 &   144 &   25.40 &  346 &   110  \nl
 25.60 &  476 &   122 &   25.60 &  319 &    96  \nl
 25.80 &  436 &   125 &   25.80 &  290 &   102  \nl
 26.00 &  419 &   126 &   26.00 &  254 &    90  \nl
 26.20 &  398 &   129 &   26.20 &  185 &    78  \nl
 26.40 &  355 &    98 &   26.40 &   81 &    43  \nl
 26.60 &  300 &   101 &   26.60 &   43 &    23  \nl
 26.80 &  257 &    86 &   26.80 &   13 &     5  \nl
 27.00 &  150 &    77 &   27.00 &    5 &     4  \nl
 27.20 &   97 &    40 &   27.20 &    0 &     1  \nl
 27.40 &   40 &    31 &   27.40 &    1 &     0  \nl
 27.60 &   14 &    11 &   27.60 &    0 &     0  \nl
 27.80 &    6 &     2 &   27.80 &    0 &     0  \nl
 28.00 &    1 &     1 &   28.00 &    0 &     0  \nl
\enddata
\end{deluxetable}

%%%%%%%%%%%%%%%%%%%%%%%%%%%%%%%%%%%%%%%%%%%%%%%%%%%%%%%%%%%%%%%%%%%%%%%%%%

\newpage

%%%%%%%%%%%%%%%%%%%%%%%%%%%%%%%%%%%%%%%%%%%%%%%%%%%%%%%%%%%%%%%%%%%%%%%%%%%


\begin{references}

\reference{AB93}
	Ashman, K. M., \& Bird, C. M.
	1993, \aj, 106, 2281

\reference{BS84}
	Bahcall, J. N., \& Soneira, R. M.  1984, \apjs, 55, 67

\reference{BV92}
	Bergbusch, P. A., \& VandenBerg, D. A.
	1992, \apjs, 81, 163

\reference{BB94}
	Bertelli, G., Bressan, A., Chiosi, C., Fagotto, F, \& Nasi, E.
	1994, \aaps, 106, 275

\reference{BB88}
	Bessell, M. S., \& Brett, J. M.
	1988, \pasp, 100, 1134

\reference{BD93}
        Bohlin, R. C., Deutsch, E. W., McQuade, K. A., Hill, J. K.,
        Landsman, W. B., O'Connell, R. W., Roberts, M. S., Smith, A. M.,
        Stecher, T. P.
        1993, \apj, 417, 127

\reference{BR95}
	Brewer, J. P., Richer, H. B., Crabtree, D. R.
	1995, \aj, 109, 2480

\reference{BH82}
	Burstein, D., \& Heiles, C.
	1982, \aj, 87, 1165

\reference{BF84}
        Burstein, D., Faber, S. M., Gaskell, C. M., \& Krumm, N.
        1984, \apj, 287, 586

\reference{BF83}
	Buzzoni, A., Fusi Pecci, F., Buonanno, R., \& Corsi, C. E.
	1983, \aap, 128, 94

\reference{CS92}
	Carney, B. W., Storm, J., \& Jones, R. V.
	1992, \apj, 386, 663

\reference{CD96}
	Chaboyer, B., Demarque, P., \& Sarajedini, A.
	1996, \apj, 459, 558

\reference{CH91}
	Christian, C. A., \& Heasley, J. N.
	1991, \aj, 101, 848

\reference{CR95}
	Couture, J., Racine, R., Harris, W. E., \& Holland, S.
	1995, \aj, 109, 2050

\reference{C86}
	Crotts, A. P. S.
	1986, \aj, 92, 292

\reference{DA90}
	Da Costa, G. S., \& Armandroff, T. E.
	1990, \aj, 100, 162

\reference{Da93}
	Davidge, T. J.
	1993, \apj, 409, 190

\reference{dV85}
	de Vaucouleurs, G.
	1958, \apj, 128, 465

\reference{DH94}
	Durrell, P. R., Harris, W. E., \& Pritchet, C. J.
	1994, \aj, 108, 2114

\reference{F95}
	Faber, S.
	1995, in Stellar Populations,
	eds.\ L. G. Gilmore, \& P. van der Kruit,
	(Dordrecht, Kluwer),
	IAU Symp.\ 169, (in press)

\reference{FM90}
	Freedman, W. L., \& Madore, B. F.
	1990, \apj, 365, 186

\reference{FP80}
        Frogel, J. A., Persson, S. E., \& Cohen, J. G.
        1980, \apj, 240, 785

\reference{GF95}
	Grillmair, C. J., Freeman, K. C., Irwin, M., \& Quinn, P. J.
	1995, \aj, 109, 2553

\reference{HH87}
	Hesser, J. E., Harris, W. E., VandenBerg, D.A., Allwright, J. W. B.,
	Shott, P., \& Stetson, P. B.
	1987, \pasp, 99, 739

\reference{Hod95}
	Hodder, P. J. C.
	1995, PhD. Thesis, UBC

\reference{H92}
	Hodge, P.
	1992, The Andromeda Galaxy,
	(Dordrecht, Kluwer)

\reference{H97}
	Holland, S.
	1997, PhD. Thesis, in preparation

\reference{HF96}
	Holland, S., Fahlman, G. G., \& Richer, H. B.
	1996, in preparation

\reference{HH95}
	Holtzman, J. A., \etal
	1995a, \pasp, 107, 156

\reference{HB95}
	Holtzman, J. A., Burrows, C. J., Casertano, S., Hester, J. J,
	Trauger, J. T., Waterson, A. M., \& Worthey, G. S.
        1995b, \pasp, 107, 1065 

\reference{HB91}
	Huchra, J. P., Brodie, J. P., \& Kent, S. M.
	1991, \apj, 370, 495

\reference{IG94}
	Ibata, R. A., Cilmore, G. G., \& Irwin, M. J.
	1994, \nat, 370, 194

\reference{K66}
	King, I. R.
	1966, \aj, 71, 64

\reference{L77}
	Lee, S.-W.
	1977, \aaps, 27, 381

\reference{LD94}
	Lee, Y.-K., Demarque, P., \& Zinn, R.
	1994, \apj, 423, 248

\reference{LR92}
	Lin, D. H., \& Richer, H. B.
	1992, \apjl, 388, L57

\reference{M63}
	Michie, R. W.
	1963, \mnras, 125, 127

\reference{M69}
	Moffat, A. F. J.
	1969, \aap, 3, 455

\reference{MK86}
	Mould, J., \& Kristian, J.
	1986, \apj, 305, 591

\reference{P83}
	Peacock, J. A.
	1983, \mnras, 202, 615

\reference{PvdB87}
	Pritchet, C. J. \& van den Bergh, S.
	1987, \apj, 316, 517

\reference{PvdB88}
	Pritchet, C. J., \& van den Bergh, S.
	1988, \apj, 331, 135

\reference{PvdB94}
	Pritchet, C. J. \& van den Bergh, S.
	1994, \aj, 107, 1730

\reference{RB85}
	Ratnatunga, K. U., \& Bahcall, J. N.
	1985, \apjs, 59, 63

\reference{RM95}
	Rich, R. M., \& Mighell, K. J.
	1995, \apj, 439, 145

\reference{RM96}
	Rich, R. M., Mighell, K. J., Freedman, W. L., \& Neill, J. D.
	1996, AJ 111, 768

\reference{RH96}
	Richer, H. B., Harris, W. E., Fahlman, G. G., Bell, R. A.,
	Bond, H. E., Hesser, J. E., Holland, S., Pryor, C., Stetson,
	P. B., VandenBerg, D. A., \& van den Bergh, S.
	1996, \apj, 463, 602

\reference{SL95}
	Sarajedini, A, \& Layden, A. C.
	1995, \aj, 109, 1086

\reference{SK68}
	Simoda, M. \& Kimura, H.
	1968, \apj, 151, 133

\reference{SH95}
	Smail, I., Hogg, D. W., Yan, L., \& Cohen, J. G.
	1995, \apjl, 449, L105

\reference{S87}
	Stetson, P. B.
	1987, \pasp, 99, 191

\reference{S94}
	Stetson, P. B.
	1994, \pasp, 106, 250

\reference{S96}
	Stetson, P. B.
	1996, private communication

\reference{T89}
        Tripicco, M. J.
        1989, \aj, 97, 735

\reference{vdB91}
	van den Bergh, S.
	1991, \pasp, 103, 1053

\reference{vdB93}
	van den Bergh, S.
	1993, \apj, 411, 178

\reference{ZW84}
	Zinn, R. \& West, M.
	1984, \apjs, 55, 45

\end{references}
\end{document}